\documentclass[pra,aps,twocolumn,nopacs,superscriptaddress,nofootinbib]{revtex4}

\usepackage[bbgreekl]{mathbbol}
\usepackage{appendix}
\usepackage[T1]{fontenc}
\usepackage[latin1]{inputenc}
\usepackage{lmodern}
\usepackage{graphicx}
\usepackage{dcolumn}
\usepackage{bm}
\usepackage{amsmath}
\usepackage{amssymb}
\usepackage{pst-all}
\usepackage{psfrag}
\usepackage{epsfig}
\usepackage{titletoc}
\usepackage{minitoc}

\usepackage{mathrsfs}
\usepackage{amsfonts}
\usepackage{mathtools}
\usepackage{color} 
\usepackage[colorlinks=true,linkcolor=blue,urlcolor=cyan,citecolor=blue]{hyperref}
\usepackage{braket}
\usepackage{commath}
\usepackage{enumitem}

\def\ii{\text{i}}  \def\ee{\text{e}}  \def\EF{{E_{\rm F}}}  \def\Iext{I_{\rm ext}}
\def\w0{{\omega_0}}  \def\wp{\omega_{\rm p}}  \def\sw{{s\omega}}  
\def\epsa{\epsilon^{\rm a}}  \def\epsb{\epsilon^{\rm b}}  \def\epsab{\epsilon^{\rm ab}}
\def\Rb{{\bf R}}  \def\rb{{\bf r}}  \def\jb{{\bf j}}  \def\Eb{{\bf E}}  \def\db{{\bf d}}  \def\eb{{\bf e}}
\def\sigone{\sigma^{(1)}}  \def\etaone{\eta^{(1)}}  \def\rhot{\tilde{\rho}}
\def\vxi{\vec{\xi}}  \def\vmu{\vec{\mu}}  \def\vth{{\vec{\theta}}}  \def\vep{\vec{\varepsilon}}

\begin{document}

\title{Nonlinear atom-plasmon interactions enabled by nanostructured graphene}

\author{Joel~D.~Cox}
\email{joel.cox@icfo.eu}
\affiliation{ICFO-Institut de Ciencies Fotoniques, The Barcelona Institute of Science and Technology, 08860 Castelldefels (Barcelona), Spain}
\author{F.~Javier~Garc\'{\i}a~de~Abajo}
\affiliation{ICFO-Institut de Ciencies Fotoniques, The Barcelona Institute of Science and Technology, 08860 Castelldefels (Barcelona), Spain}
\affiliation{ICREA-Instituci\'o Catalana de Recerca i Estudis Avan\c{c}ats, Passeig Llu\'{\i}s Companys 23, 08010 Barcelona, Spain}

\begin{abstract}
Electrically tunable graphene plasmons are anticipated to enable strong light-matter interactions with resonant quantum emitters. However, plasmon resonances in graphene are typically limited to infrared frequencies, below those of optical excitations in robust quantum light sources and many biologically interesting molecules. Here we propose to utilize near fields generated by the plasmon-assisted nonlinear optical response of nanostructured graphene to resonantly couple with proximal quantum emitters operating in the near-infrared. We show that the nonlinear near-field produced by a graphene nanodisk can strongly excite and coherently control quantum states in two- and three-level atomic systems when the third harmonic of its plasmon resonance is tuned to a particular electronic transition. In the present scheme, emitter and plasmon resonances are nondegenerate, circumventing strong enhancement of spontaneous emission. We envision potential applications for the proposed nonlinear plasmonic coupling scheme in sensing and temporal quantum control.
\end{abstract}

\pacs{73.20.Mf, 78.67.Wj, 42.65.-k}
\maketitle
\tableofcontents



\section{Introduction}

Precise control over light-matter interactions on nanometric length scales presents opportunities in technologies such as light energy harvesting \cite{AP10}, optical biosensing \cite{DSW16}, and quantum information \cite{LMS15}. Plasmons, consisting of electromagnetic fields hybridized with collective charge-carrier oscillations, are ideal for this purpose, as they couple strongly with impinging light and can focus it into nanoscale volumes \cite{M07,SBC10}. The interaction of enhanced local fields produced by resonantly illuminated plasmonic nanostructures with quantum emitters (QEs), such as atoms or molecules (natural or artificial), is of particular interest to the quantum optics and optical sensing communities \cite{A10,B12,K17}. Typically, optimal coupling is achieved by engineering the geometry of a noble metal nanostructure so that a plasmon resonates with an electronic transition in a proximal dipole emitter (e.g., an exciton in a quantum dot). However, because noble metal plasmons cannot easily be modified using external stimuli \cite{BSA15}, resonant light-matter interactions in noble metal-QE nanocomposites typically lack the active tunability required by many photonic device functionalities, and are further hindered by large intrinsic Ohmic losses \cite{K15_2}.

In its pristine form, graphene, the atomically thin carbon layer, is a zero-gap semiconductor that presents broadband 2.3\% optical absorption from transitions between linear valence and conduction bands \cite{CGP09}. When doped to a Fermi energy $\EF$, Pauli blocking prohibits vertical interband absorption for photon energies $\hbar\omega\leq2\EF$, and graphene plasmons, consisting of coherently-coupled virtual intraband transitions, emerge within this gap. These resonances exhibit stronger confinement and longer lifetimes than their noble metal counterparts, and are readily tuned in an active manner via electrostatic gating \cite{GPN12}. For these reasons, the interaction of plasmonic near fields from highly doped graphene with proximal QEs has been predicted to enable observable vacuum Rabi splittings \cite{paper176}, large Purcell enhancement factors \cite{NGG11_2,SRS18}, and electrical control of quantum states \cite{paper204}. Unfortunately, even at high doping levels, plasmon energies $\hbar\wp$ in graphene nanostructures appear in the infrared or terahertz regimes, well-below the frequencies associated with long-lived electronic excitations in robust quantum light sources. Indeed, graphene structures with a characteristic size $D$ exhibit localized plasmons at energies scaling like $\hbar\wp\propto\sqrt{\EF/D}$, which for achievable values of $\EF<1$\,eV and $D>10$\,nm have so far been demonstrated at mid-infrared and lower frequencies. Additionally, coupling to optical phonons persists as a strong source of damping for plasmons at energies above $\sim0.2$\,eV \cite{YLZ12}, as do interband transitions unless $\hbar\wp\lesssim\EF$ \cite{paper235}.

\begin{figure*}[t]
\includegraphics[width=1\textwidth]{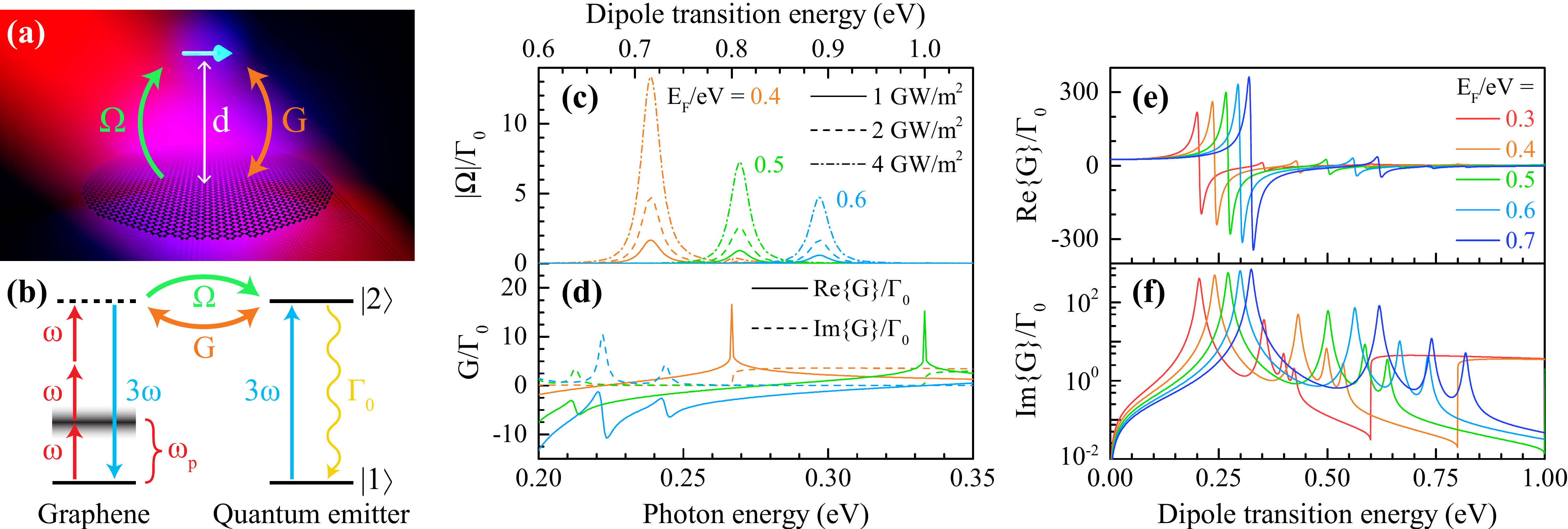}
\caption{{\bf Nonlinear atom-plasmon near-field coupling.} {\bf (a)} Illustration of a graphene nanodisk interacting with impinging light at frequency $\omega$ (red beam) by generating a near-field at $3\omega$ (blue field) that couples with a point-dipole quantum emitter (QE) located a distance $d$ above the disk center. Direct coupling of the third-harmonic (TH) near field with the QE is quantified by a Rabi frequency $\Omega$, while $G$ characterizes the QE self-interaction enabled by graphene. {\bf (b)} Energy-level diagram of the graphene-QE hybrid: Light at frequency $\omega\approx\wp$ drives plasmon-assisted TH generation in the nanodisk, producing a field at $3\omega$ that couples with a QE electronic transition of similar energy. For a $D=40$\,nm-diameter nanodisk with a QE placed $d=20$\,nm above its center and cw light impinging normally to the graphene plane, we plot {\bf (c)} $\Omega$ for various light intensities and {\bf (d)} $G$, as functions of optical frequency, with Fermi energies $\EF$ indicated by the color-coordinated curves. For the same system, the {\bf (e)} real and {\bf (f)} imaginary parts of $G$ are plotted over a wider range of emitter resonance energies.}
\label{OG}
\end{figure*}

The conical electronic dispersion of graphene imposes anharmonic intraband charge-carrier motion, resulting in an intrinsically nonlinear response to external electromagnetic fields, including efficient generation of odd-ordered harmonics \cite{M07_2,I10,CVS15,M16,JHC18,SWR18,HKD18}. The synergistic combination of a large optical nonlinearity and strong plasmonic near-field enhancement, both arising from intraband transitions in doped graphene, is currently motivating intensive research efforts in the emerging field of nonlinear graphene plasmonics \cite{paper247,JC15,CYJ15,CHC16,JKW16,paper287,paper293,KDM18}. In particular, 2D plasmons in graphene nanostructures have been predicted to generate harmonics with efficiency above that of similarly sized noble metal nanostructures \cite{paper247}. 

In this work, we propose to harness plasmon-enhanced harmonic generation in nanostructured graphene to couple the 2D material with a nearby QE, effectively bridging the energy mismatch between the electrically tunable infrared graphene plasmon and a near-infrared excitation in the emitter. We show that resonant excitation of a localized graphene plasmon can reduce the impinging light intensity required to generate a substantial third-harmonic (TH) near field, which can be actively tuned to the desired emitter resonance frequency by exploiting the unique electro-optic response of graphene. For the realistic parameters considered in this work, our simulations predict strong nonlinear plasmonic interactions that can drive fluorescence in two-level atoms and electromagnetically induced transparency (EIT) or coherent population control in three-level QEs. The present nonlinear coupling strategy can be used generally to probe nonlinear plasmonic near fields \cite{MHG17,YBY17}, while the electrical tunability of graphene plasmons can be exploited to control single-photon emission and actively manipulate quantum states.

\section{Theoretical model}

A prototypical hybrid molecule consists of a doped graphene nanodisk with diameter $D=40$\,nm and a two-level QE located a distance $d$ directly above its center, the latter characterized by the dipole moment $\vmu_{12}$ associated with the transition between states $\ket{1}$ and $\ket{2}$, which we consider to be oriented parallel to the graphene plane with magnitude $1\,e\times$\,nm ($\sim50$ Debye, commensurate with quantum dot excitons \cite{SLS01}). We consider local, isotropic dielectric media above and below the graphene layer characterized by the permittivities $\epsa_\omega$ and $\epsb_\omega$, respectively. As illustrated schematically in Fig.\ \ref{OG}(a), an impinging light electric field $\Eb_{\rm ext}$ with frequency $\omega\approx\wp$ (red beam) generates a plasmon-enhanced TH near-field, $\Eb^{33}_{\rm ind}(\rb,\omega)\ee^{-\ii3\omega t}+{\rm c.c.}$, around the nanodisk (blue field), which we describe semianalytically by adopting a classical electrostatic eigenmode decomposition (see Ref.\ \cite{paper293} and Appendix). Electron dynamics in the QE is governed by the density matrix master equation,
\begin{equation}\label{rho_eom}
\frac{\partial\rho}{\partial t}=-\frac{\ii}{\hbar}[\mathcal{H},\rho]+\mathcal{L}[\rho],
\end{equation}
where $\mathcal{H}$ is the system Hamiltonian and $\mathcal{L}[\rho]$ denotes the Lindblad superoperator accounting for decoherence. For a two-level system we write the Hamiltonian
\begin{equation}\label{H_eq}
\mathcal{H}=\hbar\sum_{j=1,2}\varepsilon_{j}\ket{j}\bra{j}-\vmu_{12}\cdot\Eb(\rb,t)\left(\ket{1}\bra{2}+\ket{2}\bra{1}\right),
\end{equation}
with $\hbar\varepsilon_j$ denoting the energy of the QE state $\ket{j}$. Defining the slowly-varying coherence elements $\rhot_{12}=\rho_{12}\ee^{\ii3\omega t}$ and $\rhot_{21}=\rho_{21}\ee^{-\ii3\omega t}$, we write the average QE transition dipole moment as $\db(t)=\vmu_{12}(\rhot_{21}\ee^{\ii3\omega t}+\rhot_{12}\ee^{-\ii3\omega t})$. Then, inserting Eq.\ (\ref{H_eq}) into Eq.\ (\ref{rho_eom}) and using the expressions for $\mathcal{L}[\rho]$ and $\Eb^{33}_{\rm ind}$ provided in the Appendix, the density matrix equations of motion in the rotating-wave approximation are found to be
\begin{align}\label{2lvl_eom}
\frac{\partial\rho_{11}}{\partial t}&=\Gamma_0\rho_{22}+\ii(\Omega^*+G^*\rhot_{12})\rhot_{21}-\ii(\Omega+G\rhot_{21})\rhot_{12}, \\
\frac{\partial\rhot_{21}}{\partial t}&=\left(\ii\Delta-\frac{\Gamma_0}{2}\right)\rhot_{21}-\ii(\Omega+G\rhot_{21})(\rho_{22}-\rho_{11}), \nonumber
\end{align}
where we adopt $\Gamma_0=1$\,ns$^{-1}$ as a phenomenological relaxation rate from the excited state $\ket{2}$ to the ground state $\ket{1}$, $\Delta\equiv3\omega-\varepsilon_{12}$ is the detuning of the TH near-field from the QE resonance energy $\hbar\varepsilon_{12}\equiv\hbar\varepsilon_2-\hbar\varepsilon_1$,
\begin{align}\label{O_eq}
\Omega=&\frac{\ii\sigma^{(3)}_{3\omega}}{3\hbar\omega\epsilon^{\rm a}_{3\omega}D}\sum_{m,m',m'',m'''}\zeta^{(3)}_{mm'm''m'''}
\frac{\vmu_{12}\cdot\eb_m}{1-\eta^{(1)}_{3\omega}/\eta_m} \\
&\times
\frac{\vxi_{m'}\cdot\Eb_{\rm ext}}{1-\etaone_\omega/\eta_{m'}}
\frac{\vxi_{m''}\cdot\Eb_{\rm ext}}{1-\etaone_\omega/\eta_{m''}}
\frac{\vxi_{m'''}\cdot\Eb_{\rm ext}}{1-\etaone_\omega/\eta_{m'''}} \nonumber
\end{align}
is the Rabi frequency quantifying the direct coupling of the QE with the TH near-field produced by the 2D nanostructure, and
\begin{equation}\label{G_eq}
G=\frac{\ii\sigma^{(1)}_{3\omega}}{3\hbar\omega(\epsilon^{\rm a}_{3\omega})^2 D^4}\sum_m\frac{\left(\vmu_{12}\cdot\eb_m\right)^2}{1-\eta^{(1)}_{3\omega}/\eta_m}
\end{equation}
characterizes the self-interaction strength of the induced QE dipole at frequency $s\omega$ enabled by graphene. In the above expressions, 
\begin{equation}\label{p_eq}
\eb_m(\rb)=\int d^2\Rb\rho_m(\Rb)\frac{\rb-\Rb}{\abs{\rb-\Rb}^3}
\end{equation}
denotes the normalized electric field produced at $\rb=(0,0,d)$ by integrating the so-called \textit{plasmon wave function} (see Ref.\ \cite{paper303} and Appendix) $\rho_m(\Rb)$ associated with mode $m$ over the 2D position vector $\Rb$ within the graphene nanostructure. The dimensionless parameters $\etaone_\sw\equiv2\ii\sigone_\sw/\sw D(\epsilon^{\rm a}_\sw+\epsilon^{\rm b}_\sw)$ in Eqs.\ (\ref{O_eq}) and (\ref{G_eq}), containing the dependence on the linear conductivities of extended graphene $\sigone_\sw$ at frequency $\sw$ (calculated in the local limit of the random-phase approximation \cite{paper235,WSS06}), yield resonant spectral features in $\Omega$ and $G$ at frequencies $\sw$ satisfying ${\rm Re}\{\etaone_\sw/\eta_m\}=1$, where $\eta_m$ are modal eigenvalues, with strengths determined by the corresponding dipolar coupling parameters $\vxi_m=\int d^2\Rb\Rb\rho_m(\Rb)/D^3$. TH generation is characterized by the local third-order conductivity of extended graphene $\sigma^{(3)}_{3\omega}$, for which we adopt the analytical result reported in Ref.\ \cite{M16}, obtained quantum-mechanically in the Dirac cone approximation, while $\zeta^{(3)}_{}$ is a third-order coupling parameter (see Appendix). The conductivities employed here, accounting for both interband and intraband electronic transitions in extended graphene at zero temperature, combined with tabulated values of $\eta_m$, $\vxi_m$, and $\zeta^{(3)}$, can faithfully reproduce the linear and nonlinear polarizabilities of graphene nanostructures with sizes $\gtrsim10$\,nm predicted in fully-atomistic simulations \cite{paper293}.

\section{Results and discussion}

Nonlinear plasmonic near-field coupling between the nanodisk and QE occurs when the transition energy $\hbar\varepsilon_{12}$ and the plasmon energy $\hbar\wp$ satisfy $\varepsilon_{12}\approx3\wp$, as indicated in Fig.\ \ref{OG}(b). The QE undergoes Rabi oscillations at a frequency $\Omega$, while the real and imaginary parts of $G$ contribute to a transition energy renormalization and decay rate enhancement, respectively \cite{ZGB06,MM11,ACM12}. Using plasmon wave functions and eigenvalues tabulated in Ref.\ \cite{paper286} for a disk geometry, we plot $\Omega$ and $G$ in Figs.\ \ref{OG}(c) and \ref{OG}(d), respectively, for a linearly polarized field parallel to $\vmu_{12}$ impinging on the QE-nanodisk system in vacuum ($\epsa_\omega=\epsb_\omega=1$). With realistic graphene parameters (Fermi energy $\EF\approx0.3$--0.7\,eV and $66$\,fs plasmon lifetime, used throughout this work), a modest separation $d=20$\,nm yields Rabi frequencies $\Omega\gtrsim\Gamma_0$, indicating strong driving of the QE via plasmon-assisted TH generation. Figures\ \ref{OG}(c) and \ref{OG}(d) show that the atom-plasmon interaction at the TH can be electrically tuned over a larger range of higher frequencies otherwise obtained though linear coupling at the fundamental frequency, while strong transition energy shifts ({\rm Re}\{G\}) and decay rate enhancement ({\rm Im}\{G\}) associated with $\wp$ (e.g., in a linear coupling scheme) are avoided at the TH, where only weakly interacting, higher-order plasmon modes play a role.

\begin{figure}[t]
\includegraphics[width=1\columnwidth]{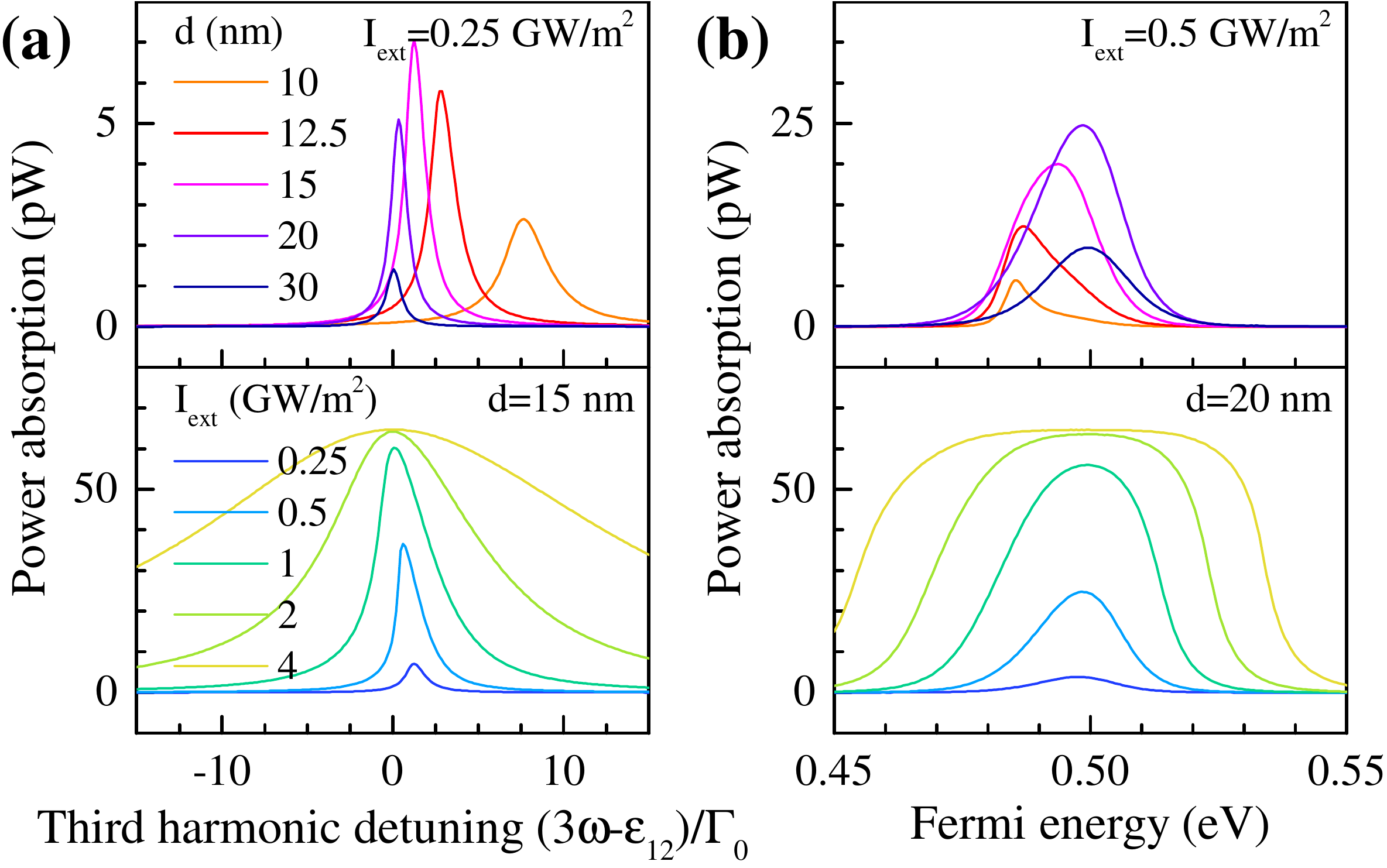}
\caption{{\bf Power absorption via third-harmonic atom-plasmon coupling.} {\bf (a)} We plot the steady-state power absorption in the QE for the system considered in Fig.\ \ref{OG} with a fixed graphene doping $\EF=0.5$\,eV (yielding a plasmon resonance $\hbar\wp=0.2694$\,eV for a disk of diameter 40\,nm), as a function of the illumination frequency when the QE resonance frequency is $\varepsilon_{12}=3\wp$. In the upper panel we plot $P_{\rm QE}$ for a fixed driving intensity $\Iext$ and various QE-graphene separations $d$, while in the lower panel we fix $d$ and vary $\Iext$. {\bf (b)} Graphene doping dependence for the system in (a) at the resonance condition $\hbar\omega=\hbar\wp=\hbar\varepsilon_{12}/3=0.2694$\,eV.}
\label{2lvl}
\end{figure}

\begin{figure}[t]
\includegraphics[width=1\columnwidth]{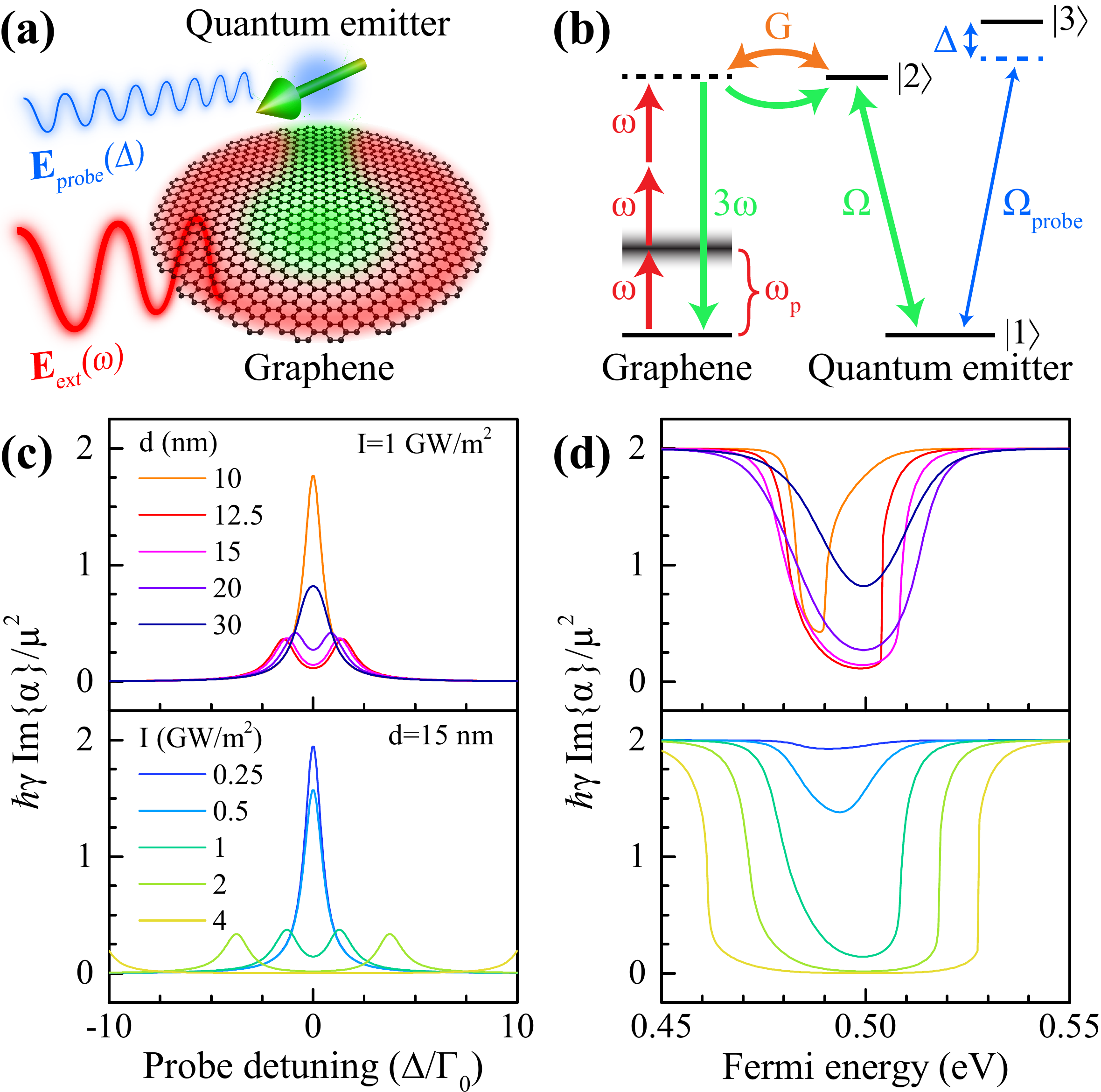}
\caption{{\bf Nonlinear near-field electromagnetically-induced transparency.} {\bf (a)} Schematic illustration of the V-type QE-graphene nanodisk hybrid system interacting with a strong driving field of frequency $\omega\approx\wp\approx\varepsilon_{12}/3$ and intensity $\Iext$ along with an auxiliary probe field of frequency $\omega_{\rm probe}$. {\bf(b)} The nonlinear plasmonic TH near-field from the nanodisk couples resonantly with the $\ket{1}\leftrightarrow\ket{2}$ transition in the QE, while the weak probe field ($\Omega_{\rm probe}=0.01\Gamma_0$ with detuning $\Delta\equiv\omega_{\rm probe}-\varepsilon_{13}$) couples only with the $\ket{1}\leftrightarrow\ket{3}$ transition. {\bf (c)} For the parameters of Fig.\ \ref{2lvl}, taking $\EF=0.5$\,eV and $\hbar\omega=\hbar\wp=\hbar\varepsilon_{12}/3=0.2694$\,eV, we plot the probe field absorption, characterized by the imaginary part of the QE polarizability associated with the $\ket{1}\leftrightarrow\ket{3}$ transition, as a function of the detuning from resonance. In the upper (lower) panel $\Iext$ ($d$) is fixed and $d$ ($\Iext$) is varied. {\bf (d)} Same as (c) but fixing $\Delta=0$ while $\EF$ is varied.}
\label{vtype}
\end{figure}

\begin{figure*}[t]
\includegraphics[width=0.85\textwidth]{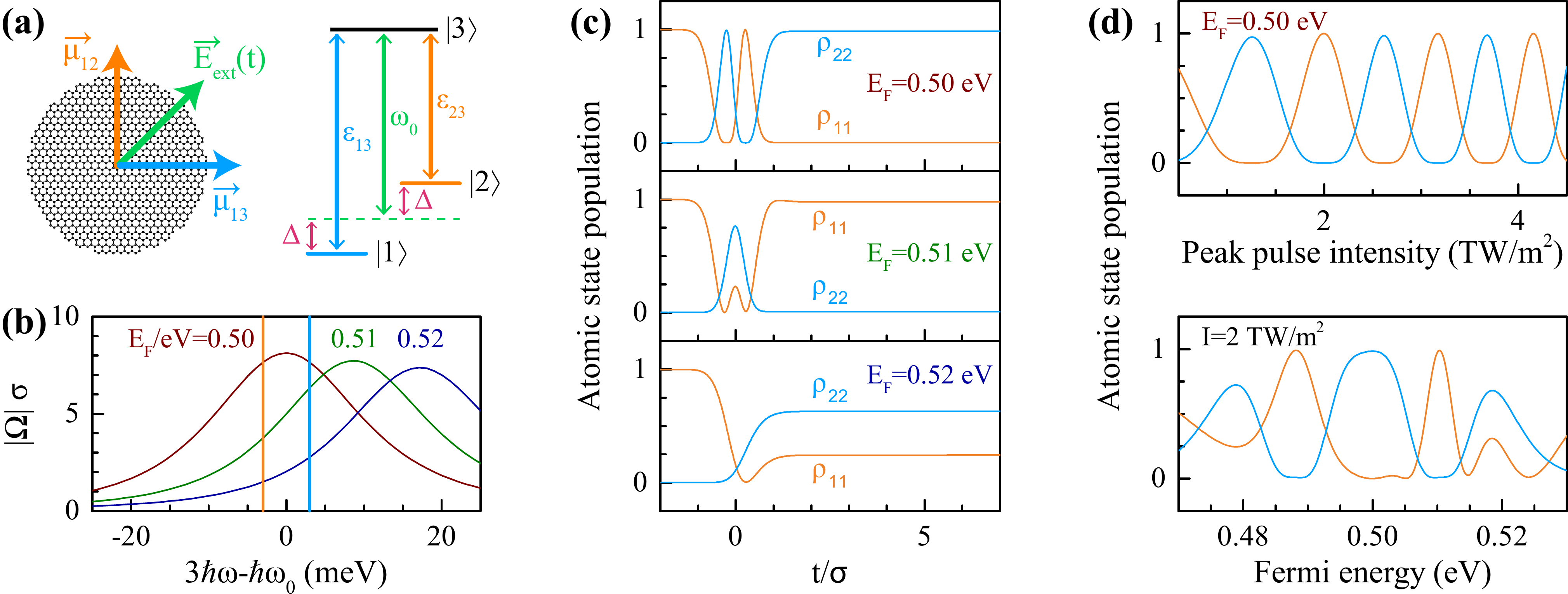}
\caption{{\bf Nonlinear plasmonic quantum control.} {\bf (a)} A schematic of the graphene nanodisk and the relative orientations of the transition dipole moments $\vmu_{12}$ and $\vmu_{13}$ of the neighboring QE, along with the incoming electric field polarization. The scheme to the right shows the energy-level structure of the $\Lambda$-type QE, indicating an equal-magnitude and opposite-sign detuning of the TH central pulse frequency from either transition. {\bf (b)} For a pulse of peak intensity $\Iext=1$\,TW/m$^2$ and FWHM duration $\sigma=130$\,fs, we plot the direct Rabi frequency at a graphene-QE separation distance $d=15$\,nm and graphene Fermi energies $\EF=0.50$, 0.51, and 0.52\,eV, with the transition energies $\varepsilon_{12}$ and $\varepsilon_{13}$ indicated by color-coordinated vertical lines. {\bf (c)} Time-evolution of the populations in states $\ket{2}$ and $\ket{3}$ for the configurations considered in (b). {\bf (d)} Final metastable state populations (at $t=10\sigma$) plotted as functions of the peak pulse intensity (upper panel) and the Fermi energy (lower panel).}
\label{lambda}
\end{figure*}

For the system considered in Fig.\ \ref{OG} under continuous-wave (cw) illumination, we simulate the QE power absorption $P_{\rm abs}=\Gamma_0\hbar\varepsilon_{12}\rho_{22}$ from the graphene plasmon TH near-field by examining the steady-state density matrix obtained upon numerical integration of Eqs.\ (\ref{2lvl_eom}). In Fig.\ \ref{2lvl}(a) we plot the TH power absorption as a function of the impinging light frequency for different QE-graphene separation distances $d$ (upper panel) and light intensities $\Iext$ (lower panel), maintaining $\EF=0.5$\,eV in graphene such that $\hbar\wp=0.2694$\,eV [see Fig.\ \ref{OG}(c)] and $\varepsilon_{12}=3\wp$. Under these conditions, the QE resonance does not interact directly with a plasmon mode of the graphene nanodisk, although a persisting self-interaction at the QE transition frequency, quantified by the parameter $G$, leads to enhanced decay rates and renormalized energies \cite{ZGB06}: this interaction is proportional to the square of the dipole coupling $\vmu_{12}\cdot\eb_m$ [see Eq.\ (\ref{G_eq})] and yields a blue-shifted and broadened power absorption peak at small separations (e.g., $d=10$\,nm) but rapidly disappears with increasing separation distance, while the direct TH near-field interaction, which depends linearly on the dipole coupling, diminishes more slowly. In the present nonlinear plasmonic coupling scheme, the TH near-field enhancement of the QE excitation rate ($\Omega$) is independent of the self-interaction strength ($G$), unlike in the degenerate plasmon-exciton coupling scheme, where these competing phenomena are both maximized at the plasmon resonance frequency \cite{DVK16,K17}. Under strong driving fields, the TH near-field produces a clear signature of power broadening in the QE absorption spectrum \cite{B08_3}, leading towards optical bistability and hysteresis fueled by the QE self-interaction \cite{MM11}. In Fig.\ \ref{2lvl}(b) we consider a situation where the illumination frequency is fixed at $\omega=\wp=\varepsilon_{12}/3$ and the TH power absorption is actively controlled by tuning the Fermi energy $\EF$ (doping charge density) in the graphene nanodisk.

The effective absorption cross-section of the hybrid system, defined as $P_{\rm abs}/I_{\rm ext}$, can reach values of approximately $\sim0.06$\,nm$^2$ (for $I_{\rm ext}=1$\,GW/m$^2$ at $d=15$\,nm), which is several orders of magnitude larger than the optimal cross-section of the dipole emitter in free space, $8\pi\mu_{12}^2\varepsilon_{12}/\hbar\Gamma_0 c\approx2.5\times10^{-5}$\,nm$^2$, but significantly less than the geometrical area of the graphene nanodisk. We further remark that although these results are obtained for intense cw illumination, approaching the damage threshold for graphene, qualitatively similar behavior is expected for pulses of similar peak intensity and durations of several nanoseconds (i.e., with energy densities falling well-below the damage threshold \cite{KLC09,CCB11,RCR11,KKK13}).

The near-field produced by plasmon-assisted TH generation can be probed in an alternative manner by QEs with more complicated energy-level structures. In a three-level V-type atom, which we study in Fig.\ \ref{vtype}, EIT \cite{FIM05} produces a dip in the absorption associated with one atomic transition when the other is coupled strongly to the TH near-field. This process is illustrated schematically in Fig.\ \ref{vtype}(a), where two external fields are applied to the hybrid system and the ground state $\ket{1}$ is coupled to two higher-energy states $\ket{2}$ and $\ket{3}$: we consider $\varepsilon_{12}=3\wp$, while the energy associated with the transition $\ket{1}\leftrightarrow\ket{3}$ is arbitrarily far away from any prominent features in the nanodisk spectrum [see Fig.\ \ref{vtype}(b)]. By probing the $\ket{1}\leftrightarrow\ket{3}$ transition with a weak field ($\Omega_{\rm probe}=0.01\Gamma_0$) while resonantly coupling the $\ket{1}\leftrightarrow\ket{2}$ transition with the TH near-field (i.e., $\omega=\wp=\varepsilon_{12}/3$), an absorption dip at resonance is observed when the QE and graphene nanodisk are in close proximity and/or the graphene plasmon is strongly driven [see Fig.\ \ref{vtype}(c)]. The active tunability of the graphene plasmon resonance can be exploited to actively switch EIT on and off, as we demonstrate in Fig.\ \ref{vtype}(d).

In Fig.\ \ref{lambda} we seek to achieve electrical control of atomic state populations using the graphene nanodisk TH near-field. In this scenario, a QE with a lambda-type energy level configuration is located above the nanodisk and has orthogonal dipole moments associated with nearly-degenerate $\ket{1}\leftrightarrow\ket{2}$ and $\ket{2}\leftrightarrow\ket{3}$ transitions [we assume $\Delta=1$\,meV; see Fig.\ \ref{lambda}(a)], which can be simultaneously driven by an optical pulse polarized $45^{\circ}$ from each of them \cite{ACM12}. We consider a Fourier-transform-limited Gaussian pulse of FWHM duration $\sigma=130$\,fs, peak intensity $I_{\rm max}$, and carrier frequency $\omega$, such that $3\omega$ is $\pm\Delta$ away from $\varepsilon_{12}$ and $\varepsilon_{13}$, respectively. The effective maximum TH Rabi frequency $\abs{\Omega}$ for cw light at the peak pulse intensity is plotted in Fig.\ \ref{lambda}(b) for various Fermi energies, with the energies $\hbar\varepsilon_{12}/3$ and $\hbar\varepsilon_{13}/3$ indicated by vertical lines. Clearly, an asymmetry in the Rabi frequency driving the two nearly degenerate transitions can be introduced by modifying the Fermi energy, resulting in different levels of population transfer from the state $\ket{2}$ to state $\ket{3}$ [Fig.\ \ref{lambda}(c,d)].

\section{Conclusions}

In conclusion, we have demonstrated that plasmon-assisted up-conversion in graphene nanostructures can be used to resonantly excite a proximal QE, enabling energy transfer, EIT, or quantum state population control that would otherwise be prevented by the energy mismatch between infrared graphene plasmons and higher-energy electronic transitions in QEs. This scheme can also be used to probe the nonlinear optical response associated with graphene plasmons, potentially alleviating the sparsity of experimental work in the field of nonlinear graphene plasmonics. While the combination of an intrinsically-large nonlinear response and intense in-plane electric field enhancement associated with localized electrically tunable plasmons renders graphene an ideal material platform for nonlinear near-field coupling, this principle could be straightforwardly applied to other plasmonic materials. We anticipate that the strategy presented here can be used to couple electrically tunable graphene plasmons with higher-energy transitions in biologically interesting molecules and high-fidelity single photon sources, opening a wide range of applications in sensing and quantum nano-optics.

\acknowledgments

The authors thank Renwen Yu for fruitful discussions and for providing the plasmon wave functions associated with a disk geometry. This work has been supported in part by the Spanish MINECO (MAT2017-88492-R and SEV2015- 0522), ERC (Advanced Grant 789104-eNANO),  the European Commission (Graphene Flagship 696656), the Catalan CERCA Program, and Fundaci\'{o} Privada Cellex.


\appendix
\appendixpage
\addappheadtotoc

\renewcommand{\thesection}{\Alph{section}}
\renewcommand{\theequation}{A\arabic{equation}}
\setcounter{section}{0}

We provide details on the semi-analytical plasmon wave function formalism used to describe the nonlinear plasmonic near-field produced by an arbitrary 2D material and its interaction with a proximal dipole emitter, summarize the theory of plasmon-assisted third-harmonic generation in a graphene nanostructure, and derive the Bloch equations governing electron dynamics in the two- and three-level atoms considered in the main text.

\section{Basic concepts and approximations}

In the main text we consider a hybrid molecule consisting of a graphene nanostructure that occupies a finite region in the $\Rb=(x,y)$ plane and a proximal quantum emitter (QE) located at $\rb=(x,y,z)$. The system is illuminated by an electromagnetic field $\Eb_{\rm ext}$ polarized in the graphene plane and oscillating at frequency $\omega\sim\wp$, where $\wp$ is the resonance frequency associated with a localized plasmon mode supported by the graphene structure. Through an $n^{\rm th}$-order nonlinear optical process, the graphene nanostructure generates a field $\Eb^{ns}_{\rm ind}(\rb)$ oscillating at a harmonic frequency $\sw$ ($\abs{s}\leq n$) that interacts with the QE. In what follows, we describe the nonlinear plasmonic near field generated by the graphene nanostructure and its interaction with the QE in a semi-analytical fashion by making the following simplifying assumptions:

\begin{itemize}[leftmargin=*]
\item{The 2D graphene nanostructure under consideration is much smaller than the light wavelength associated with $\wp$, so that we can safely describe its response in the electrostatic limit.}
\item{The QE is characterized by a point dipole placed at $\rb=(x,y,z)$ with moment $\db_\sw$ and resonance frequency $\sim\sw$ associated with the transition between two discrete electronic states.}
\item{The graphene optical response for each order $n$ and frequency $\sw$ is characterized by an isotropic conductivity $\sigma^{(n)}_\sw$ in the local limit, which is a reasonable approximation when the in-plane light momenta are much smaller than the involved electron momenta. This approximation is well justified for normally-impinging light interacting with localized plasmon modes in deeply-subwavelength 2D structures.}
\item{The impinging light is sufficiently weak so that the optical response of graphene is well described by perturbation theory, maintaining zero electron temperature.}
\item{The direct interaction of the external field with the QE is negligible compared to the near field produced by the nonlinear response of the graphene nanostructure.}
\item{Inelastic scattering processes in the graphene nanostructure are frequency-independent, and occur at a phenomenological rate $\hbar\tau^{-1}=10$\,meV ($\tau\approx66$\,fs) corresponding to a conservative electron mobility of $\sim1300$\,cm$^2$\,V$^{-1}$\,s$^{-1}$ at $\EF\approx0.5$\,eV.}
\end{itemize}

\section{Electrostatic description of nonlinear plasmonic near-fields in 2D materials}

In this section, we summarize the formalism presented in Ref.\ \cite{paper293}, which is used in this work to describe the nonlinear response of nanostructured graphene. To $n^{\rm th}$-order, an isotropic 2D material responds to an in-plane external field $\Eb_{\rm ext}\ee^{-\ii\omega t}+{\rm c.c.}$ by generating an induced charge $\rho^{ns}_{\rm ind}$ oscillating at a harmonic frequency $\sw$ ($\abs{s}\leq n$) associated with the nonlinear 2D current $\jb^{ns}$ through the continuity equation,
\begin{equation}\label{cont}
\rho^{ns}_{\rm ind}(\Rb,\omega)=-\frac{\ii}{\sw}\nabla_\Rb\cdot\jb^{ns}(\Rb,\omega).
\end{equation}
We express $\jb^{ns}$ self-consistently as the sum of currents produced by the \textit{direct} nonlinear response to the total electric field at $\omega$ ($\jb^{ns}_{\rm NL}$) and the linear response to the induced electric field at the generated frequency $\sw$ ($\Eb^{ns}$), according to
\begin{equation}\label{j_R}
\jb^{ns}(\Rb,\omega)=f_\Rb\left[\jb^{ns}_{\rm NL}(\Rb,\omega)+\sigma^{(1)}_\sw\Eb^{ns}(\Rb,\omega)\right],
\end{equation}
where $f_\Rb=1$ when $\Rb$ lies within the 2D material and is 0 everywhere else, effectively defining the structure morphology. For a QE with a resonance frequency $\sim\sw$ positioned at a point $\rb$ above the 2D material, the total electric field oscillating at this frequency within the 2D material becomes
\begin{align}\label{E_R}
\Eb^{ns}(\Rb,\omega)=-\nabla_\Rb&\left[\frac{1}{\epsa_\sw}\frac{\left(\Rb-\rb\right)\cdot\db_\sw}{\abs{\Rb-\rb}^3}\right.  \\
&\left.+\frac{1}{\epsab_\sw}\int d^2\Rb'\,\frac{\rho^{ns}_{\rm ind}(\Rb',\omega)}{\abs{\Rb-\Rb'}}\right],  \nonumber
\end{align}
with the first and second terms arising from the QE transition dipole moment $\db_\sw$ and the induced charge, respectively. Here, $\epsab_\omega=(\epsa_\omega+\epsb_\omega)/2$ denotes the average of the dielectric functions for media above ($\epsa_\omega$) and below ($\epsb_\omega$) the 2D layer.

\begin{figure*}[t]
\includegraphics[width=0.75\textwidth]{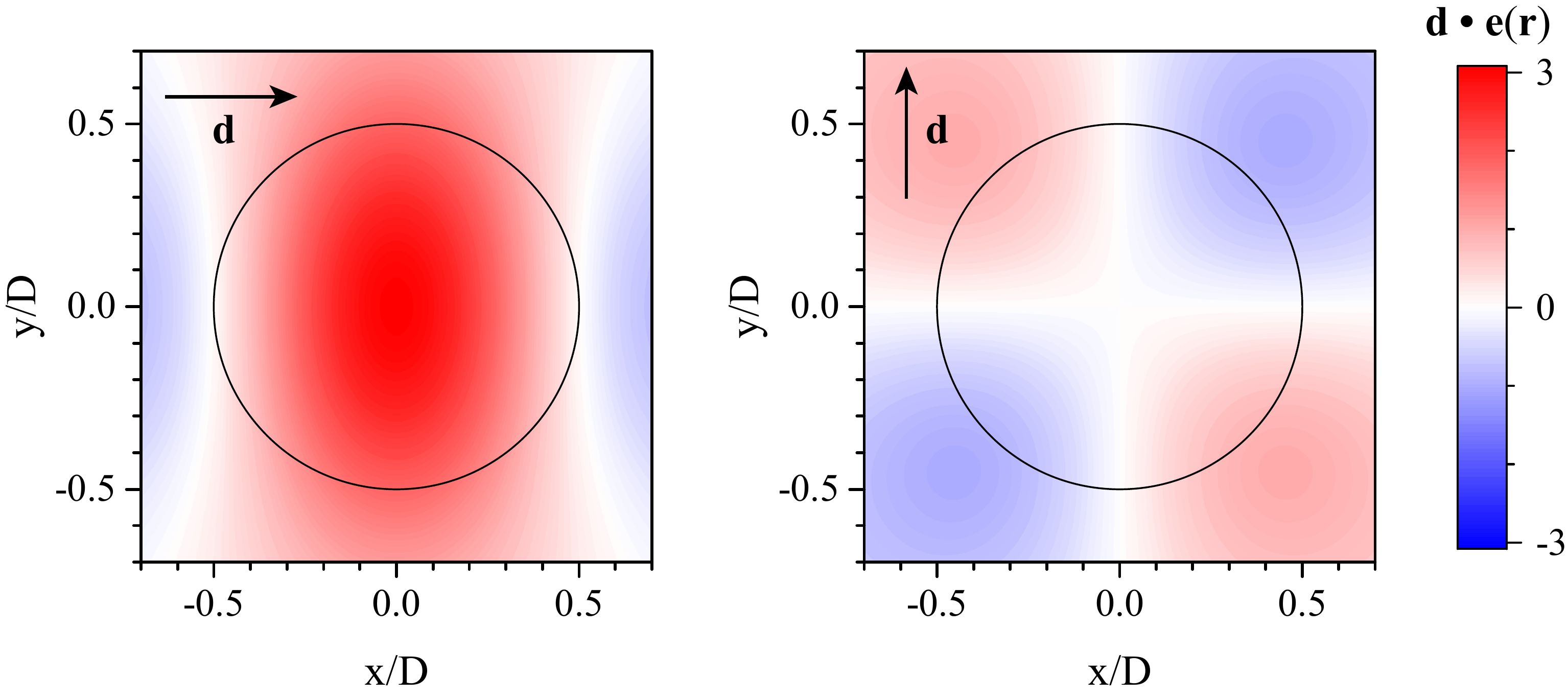}
\caption{{\bf Spatial distribution of the plasmonic near-field.} We plot the scalar product $\db\cdot\eb(\rb)$ produced by the lowest-order dipolar mode of a 2D disk with diameter $D$ (edge indicated by the black circle) in the $\Rb=(x,y)$ plane for a unit dipole $\db$ positioned in the $\rb=(x,y,0.5D)$ plane and oriented either along the $x$ direction (left panel) or $y$ direction (right panel). The impinging electric field is polarized along the $x$ direction and the disk lies in the $x$-$y$ plane.}
\label{ddote}
\end{figure*}

Now, introducing the normalized 2D coordinate vector $\vth\equiv\Rb/D$, where $D$ is the characteristic distance in the 2D geometry (e.g., the diameter of a disk or the side length of an equilateral triangle), we perform algebraic manipulations of Eqs.\ (\ref{cont}-\ref{E_R}) to write the self-consistent current $\tilde{\jb}^{ns}\equiv\jb^{ns}/\sqrt{f_\vth}$ as
\begin{align}\label{j_th}
\tilde{\jb}^{ns}(\vth,\omega)=&\sqrt{f_\vth}\,\jb^{ns}_{\rm NL}(\vth,\omega)  \\ 
&-\frac{\sigma^{(1)}_\sw}{\epsa_\sw D^3} \sqrt{f_\vth}\,\nabla_\vth\ \frac{(\vth-\rb/D)\cdot\db_\sw}{\bigl\lvert\vth-\rb/D\bigl\rvert^3}  \nonumber \\
&+\eta^{(1)}_\sw\int d^2\vth'\,{\bf M}(\vth,\vth')\cdot\tilde{\jb}^{ns}(\vth',\omega),  \nonumber
\end{align}
where $\eta^{(1)}_\sw=\ii\sigma^{(1)}_\sw/\sw\epsab_\sw D$ is a dimensionless parameter and
\begin{equation}\label{M_th}
{\bf M}(\vth,\vth')=\sqrt{f_\vth f_{\vth'}}\nabla_\vth\otimes\nabla_\vth\,\frac{1}{\bigl\lvert\vth-\vth'\bigl\rvert}.
\end{equation}
We identify ${\bf M}(\vth,\vth')$ as a real and symmetric operator associated with the 2D geometry defined by $f_\vth$ that admits real eigenmodes $\vep_m(\vth)$ and eigenvalues $1/\eta_m$ satisfying
\begin{equation}\label{eigen}
\int d^2\vth'\,{\bf M}(\vth,\vth')\cdot\vep_m(\vth')=\frac{1}{\eta_m}\vep_m(\vth)
\end{equation}
and forming an orthonormal basis, that is,
\begin{equation}\label{ortho}
\int d^2\vth\,\vep_m(\vth)\vep_{m'}(\vth)=\delta_{mm'}.
\end{equation}
Following the procedure of Ref.\ \cite{paper293}, we expand the solution to Eq.\ (\ref{j_th}) in a sum of the eigenmodes of ${\bf M}(\vth,\vth')$ according to
\begin{equation}\label{j_soln}
\tilde{\jb}^{ns}(\vth,\omega)=\sum_m a^{ns}_m(\omega)\vep_m(\vth).
\end{equation}
Inserting Eq.\ (\ref{j_soln}) into Eq.\ (\ref{j_th}), multiplying the resulting expression by $\vep_{m'}(\vth)$, and integrating over the 2D coordinates $\vth$, we make use of Eqs.\ (\ref{eigen}) and (\ref{ortho}) to isolate the expansion coefficients
\begin{equation}\label{a_ns}
a^{ns}_m(\omega)=\frac{1}{1-\eta^{(1)}_\sw/\eta_m}\left[b^{ns}_m(\omega)+\frac{\sigma^{(1)}_\sw}{\epsa_\sw D^3}\db_\sw\cdot\eb_m\right],
\end{equation}
where we have introduced the nonlinear coupling coefficient
\begin{equation}\label{b_ns}
b^{ns}_m(\omega)=\int d^2\vth\,\sqrt{f_\vth}\,\vep_m(\vth)\cdot\jb^{ns}_{\rm NL}(\vth,\omega)
\end{equation}
and the normalized electric field generated by the $m^{\rm th}$ eigenmode of the 2D geometry
\begin{equation}\label{p_m}
\eb_m(\rb)=\int d^2\vth\,\frac{\vth-\rb/D}{\bigl\lvert\vth-\rb/D\bigr\rvert^3}\rho_m(\vth)
\end{equation}
in terms of the so-called \textit{plasmon wave function} (PWF) associated with it, $\rho_m(\vth)\equiv\nabla_\vth\cdot\sqrt{f_\vth}\,\vep_m(\vth)$ \cite{paper303}. In terms of normalized coordinates, Eq.\ (\ref{cont}) provides the induced charge density $\rho^{ns}_{\rm ind}(\vth,\omega)=-(\ii/\sw D)\sum_m a^{ns}_m \rho_m(\vth)$, from which we use Eq.\ (\ref{a_ns}) to express the electric field acting on the QE at $\rb$ as
\begin{align}\label{E_ns}
\Eb^{ns}_{\rm ind}(\rb,\omega)=&\frac{\ii}{\sw\epsa_\sw D}\sum_m\frac{\eb_m(\rb)}{1-\eta^{(1)}/\eta_m}  \\
&\times \left[b^{ns}_m(\omega)+\frac{\sigma^{(1)}_\sw}{\epsa_\sw D^3}\db_\sw\cdot\eb_m(\rb)\right].  \nonumber
\end{align}
The first term in the above expression describes the near field produced by the 2D material oscillating at frequency $\sw$ and generated directly by the external field at $\omega$, while the second term accounts for the field produced by the QE dipole oscillating at $\sw$. In practice, the PWFs $\rho_m(\vth)$, eigenmodes $\vep_m(\vth)$, and eigenvalues $\eta_m$ of a given structure morphology are numerically computed once and for all using an electrostatic solver. In the main text, the PWFs and eigenvalues of a circular nanodisk are obtained from tabulated data reported in Ref.\ \cite{paper286}. From Eq.\ (\ref{E_ns}) we conclude that the coupling of the QE dipole $\db_{s\omega}$ with the TH near-field produced by the $m^{\rm th}$ mode of a 2D geometry is proportional to the scalar product $\db_{s\omega}\cdot\eb_m(\rb)$, which is strongly dependent on the position of the dipole. In Fig.\ \ref{ddote}, we study this coupling for a dipole placed at a fixed distance away from the plane occupied by a circular disk and oriented either parallel or perpendicular to the polarization of the incident electric field. Evidently, maximal coupling should occur for a dipole aligned with the external field polarization and centered on the disk.

The 2D nanostructure response to a monochromatic external field $\Eb_{\rm ext}(\Rb,\omega)$ is dominated by the \textit{linear} self-consistent field (i.e., $n=s=1$),
\begin{align}
\Eb^{11}(\Rb,\omega)=&\Eb_{\rm ext}(\Rb,\omega)  \\
&-\frac{1}{\epsab_\omega}\nabla_\Rb\int d^2\Rb'\,\frac{\rho^{11}_{\rm ind}(\Rb',\omega)}{\abs{\Rb-\Rb'}},  \nonumber
\end{align}
where $\rho^{11}_{\rm ind}$ is the induced charge given by Eq.\ (\ref{cont}). Following Ref.\ \cite{paper303}, we write the linear current as $\jb^{11}=\sigma^{(1)}_{\omega}\Eb^{11}$, move to $\vth$ coordinates, and define the normalized field $\vep\,(\vth,\omega)\equiv D\sqrt{f_\vth}\,\Eb^{11}(\vth,\omega)$ to write
\begin{equation}
\vep\,(\vth,\omega)=\vep_{\rm ext}(\vth,\omega)+\eta^{(1)}_\omega\int d^2\vth'\,{\bf M}(\vth,\vth')\cdot\vep\,(\vth',\omega).
\end{equation}
Expanding the solution to the above expression in the eigenmodes of ${\bf M}(\vth,\vth')$ [see Eqs.\ (\ref{M_th}-\ref{ortho})], we obtain the normalized field
\begin{equation}\label{E11_th}
\vep\,(\vth,\omega)=\sum_m\frac{c_m}{1-\eta^{(1)}_\omega/\eta_m}\vep_m(\vth),
\end{equation}
where the expansion coefficients are
\begin{equation}
c_m=\int d^2\vth\,\vep_m(\vth)\cdot\vep_{\rm ext}(\vth,\omega).
\end{equation}
For plane-wave illumination normal to the graphene plane, $\Eb_{\rm ext}(\Rb,\omega)$ is independent of $\Rb$ and Eq.\ (\ref{E11_th}) yields the electric field
\begin{equation}\label{E11}
\Eb^{11}(\vth,\omega)=\sum_m\frac{-\vxi_m\cdot\Eb_{\rm ext}}{1-\eta^{(1)}_\omega/\eta_m}\frac{\vep\,(\vth)}{\sqrt{f_\vth}},
\end{equation}
where the dimensionless parameter $\vxi_m=\int d^2\vth\,\vth\,\rho_m(\vth)$ 
characterizes the strength of the induced dipole \cite{paper303}. The in-plane near-fields given by Eq.\ (\ref{E11}) depend on the linear conductivity $\sigma^{(1)}_\omega$ of the 2D material entering the parameter $\eta^{(1)}_\omega$. For graphene doped to a Fermi energy $\EF$ and an inelastic electron scattering time $\tau$, we adopt the local conductivity obtained in the random-phase approximation at zero temperature,
\begin{align}
\sigma^{(1)}_\omega=&\frac{\ii e^2}{\pi\hbar^2}\frac{\EF}{\omega+\ii\tau^{-1}}  \\
&+\frac{e^2}{4\hbar}\left[\Theta(\hbar\omega-2\EF)+\frac{\ii}{\pi}\log\abs{\frac{\hbar\omega-2\EF}{\hbar\omega+2\EF}}\right],  \nonumber
\end{align}
with the first and second terms accounting for intra- and inter-band optical transitions, respectively \cite{WSS06}.

The plasmon-enhanced in-plane field at frequency $\omega$, given to linear order by Eq.\ (\ref{E11_th}), drives the nonlinear current $\jb^{ns}_{\rm NL}$ entering Eq.\ (\ref{b_ns}), producing the near-field described by Eq.\ (\ref{E_ns}). For third-harmonic generation (i.e., $n=s=3$), the nonlinear source current is
\begin{equation}\label{j33}
\jb^{33}_{\rm NL}=\sigma^{(3)}_{3\omega}\left(\Eb^{11}\cdot\Eb^{11}\right)\Eb^{11},
\end{equation}
where $\sigma^{(3)}_{3\omega}$ is the third-order conductivity characterizing third-harmonic generation in the infinitely-extended 2D material. Here, we adopt the analytical result reported in Ref.\ \cite{M16}, which is obtained from a local quantum-mechanical description of graphene in the Dirac cone approximation at zero temperature. Inserting Eq.\ (\ref{E11}) into Eq.\ (\ref{j33}), we thus obtain from Eq.\ (\ref{b_ns})
\begin{align}
b^{33}_m(\omega)=&-\sigma^{(3)}_{3\omega}\sum_{m'm''m'''}\zeta^{(3)}_{mm'm''m'''}  \\
&\times\frac{\vxi_{m'}\cdot\Eb_{\rm ext}}{1-\eta^{(1)}_\omega/\eta_{m'}}\frac{\vxi_{m''}\cdot\Eb_{\rm ext}}{1-\eta^{(1)}_\omega/\eta_{m''}}\frac{\vxi_{m'''}\cdot\Eb_{\rm ext}}{1-\eta^{(1)}_\omega/\eta_{m'''}},  \nonumber
\end{align}
where
\begin{equation}
\zeta^{(3)}_{mm'm''m'''}=\int d^2\vth\,\frac{1}{f_\vth}\vep_m(\vth)\cdot\vep_{m'}(\vth)\vep_{m''}(\vth)\cdot\vep_{m'''}(\vth)
\end{equation}
is a unitless parameter. Mode dipoles $\vxi_m$ and eigenvalues $\eta_m$ of a given structure morphology are numerically computed once and for all using an electrostatic solver. Throughout the main text we consider a circular graphene nanodisk, for which the linear $\vxi_m$ and nonlinear $\zeta^{(3)}_{mm'm''m'''}$ dipolar coupling parameters for modes up to order $m\leq3$ have been tabulated in Ref.\ \cite{paper293}.

\section{Nonlinear 2D plasmon--quantum emitter interaction}

Electron dynamics in the two- and three-level quantum emitters (QEs) considered in the main text is governed by the Lindblad master equation for the time-evolution of the density matrix,
\begin{equation}\label{master}
\dot{\rho}=-\frac{\ii}{\hbar}[\mathcal{H},\rho]+\mathcal{L}[\rho],
\end{equation}
where $\mathcal{H}$ is the system Hamiltonian and $\mathcal{L}[\rho]$ denotes the Lindblad superoperator accounting for decoherence. In what follows, we consider that the QE states $\ket{i}$ with energies $\hbar\varepsilon_i$ form a complete and orthonormal set. This allows us to express the density matrix as $\rho=\sum_{ij}\rho_{ij}\sigma_{ij}$, where $\sigma_{ij}\equiv\ket{i}\bra{j}$, noting that the condition $\rho_{ij}=\rho^*_{ji}$ is guaranteed by the Hermitian property $\rho^\dagger=\rho$. Energies and dipole moments associated with transitions $\ket{i}\leftrightarrow\ket{j}$ are denoted by $\varepsilon_{ij}\equiv\varepsilon_j-\varepsilon_i$ and $\vmu_{ij}$, respectively. For simplicity we assume that all incoherent decay processes occur at the same phenomenological rate $\Gamma_0$.

\subsection{Two-level atom}

For a two-level QE with a ground state $\ket{1}$ and excited state $\ket{2}$ interacting with the classical monochromatic field $\Eb(\rb,t)=\Eb^{ns}_{\rm ind}(\rb,\omega)\ee^{-\ii\sw t}+{\rm c.c.}$, where $\sw\sim\varepsilon_{12}$, we insert the Hamiltonian
\begin{equation}\label{H_QE}
\mathcal{H}=\hbar\sum^2_{j=1} \varepsilon_i\sigma_{jj}-\vmu_{12}\cdot\Eb(\rb,t)\left(\sigma_{12}+\sigma_{21}\right)
\end{equation}
and the Lindblad operator
\begin{equation}
\mathcal{L}[\rho]=\frac{\Gamma_0}{2}\left(2\sigma_{12}\rho\sigma_{21}-\sigma_{21}\sigma_{12}\rho-\rho\sigma_{21}\sigma_{12}\right)
\end{equation}
into Eq.\ (\ref{master}). Making use of Eq.\ (\ref{E_ns}) and writing the QE dipole moment as $\db(t)=\vmu_{12}(\tilde{\rho}_{21}\ee^{-\ii\sw t}+\tilde{\rho}_{12}\ee^{\ii\sw t})$, where $\tilde{\rho}_{21}=\tilde{\rho}^*_{12}=\rho_{21}\ee^{\ii\sw t}$, the density matrix equations of motion in the rotating-wave approximation are obtained as
\begin{subequations}
\begin{align}\label{eom2lvl}
\frac{\partial\rho_{11}}{\partial t}&=\Gamma_0\rho_{22}+\ii\left(\Omega^*+G^*\tilde{\rho}_{12}\right)\tilde{\rho}_{21}-\ii\left(\Omega+G\tilde{\rho}_{21}\right)\tilde{\rho}_{12}, \\
\frac{\partial\tilde{\rho}_{21}}{\partial t}&=\left(\ii\Delta-\frac{\Gamma_0}{2}\right)\tilde{\rho}_{21}-\ii\left(\Omega+G\tilde{\rho}_{21}\right)\left(\rho_{22}-\rho_{11}\right),
\end{align}
\end{subequations}
where $\Delta\equiv\sw-\varepsilon_{12}$ is the detuning parameter, 
\begin{equation}\label{O}
\Omega=\frac{\ii}{s\hbar\omega\epsa_\sw D}\sum_m b^{ns}_m(\omega)\frac{\vmu_{12}\cdot\eb_m}{1-\eta^{(1)}_\sw/\eta_m}
\end{equation}
is the Rabi frequency, which describes the direct coupling between the QE and the nonlinear near-field produced in the 2D material at frequency $\sw$, and
\begin{equation}\label{G}
G=\frac{\ii\sigma^{(1)}_\sw}{s\hbar\omega(\epsa_\sw)^2 D^4}\sum_m\frac{\left(\vmu_{12}\cdot\eb_m\right)^2}{1-\eta^{(1)}_\sw/\eta_m}  
\end{equation}
quantifies the self-interaction of the induced QE dipole at $s\omega$ enabled by the 2D nanostructure.

\subsection{Three-level V-type atom: Electromagnetically induced transparency}

For a V-type atom, we consider a QE with ground state $\ket{1}$ and excited states $\ket{2}$ and $\ket{3}$ interacting with the classical light field
\begin{equation}
\Eb(\rb,t)=\Eb^{ns}_{\rm ind}(\rb,\omega)\ee^{-\ii\sw t}+\Eb_{\rm probe}\ee^{-\ii\omega_{\rm probe}t}+{\rm c.c.},
\end{equation} 
where $\sw\sim\varepsilon_{12}$ and $\omega_{\rm probe}\sim\varepsilon_{13}$, but $\abs{\varepsilon_2-\varepsilon_3}\gg\Gamma_0$ (i.e., levels $\ket{2}$ and $\ket{3}$ are spaced far away from one another in energy so that only $\Eb^{ns}_{\rm ind}$ couples with $\vmu_{12}$ and only $\Eb_{\rm probe}$ couples with $\vmu_{13}$). In the rotating-wave approximation, the Hamiltonian describing this system is
\begin{equation}
\mathcal{H}=\hbar\sum^3_{i=1}\varepsilon_i\sigma_{ii}-\left[\begin{array}{c}\vmu_{12}\cdot\Eb^{ns}_{\rm ind}(\rb,\omega)\ee^{-\ii\sw t}\sigma_{21}  \\+\vmu_{13}\cdot\Eb_{\rm probe}\ee^{-\ii\omega_{\rm probe} t}\sigma_{31}  \\ +{\rm h.c.}\end{array}\right],
\end{equation}
and the Lindblad operator is
\begin{equation}
\mathcal{L}[\rho]=\sum^3_{j=2}\frac{\Gamma_j}{2}\left(2\sigma_{1j}\rho\sigma_{j1}-\sigma_{j1}\sigma_{1j}\rho-\rho\sigma_{j1}\sigma_{1j}\right),
\end{equation}
where $\Gamma_j$ is the decay rate of state $\ket{j}$ to $\ket{1}$. Inserting the above expressions into Eq.\ (\ref{master}), we obtain the equations of motion for the density matrix elements:
\begin{subequations}
\begin{align}
\frac{\partial\rho_{22}}{\partial t}=&-\Gamma_2\rho_{22}+\ii\left(\Omega_2+G\tilde{\rho}_{21}\right)\tilde{\rho}_{12}  \\
&-\ii\left(\Omega_2^*+G^*\tilde{\rho}_{12}\right)\tilde{\rho}_{21}, \nonumber \\
\frac{\partial\rho_{33}}{\partial t}=&-\Gamma_3\rho_{33}+\ii\Omega_3\tilde{\rho}_{13}-\ii\Omega^*_3\tilde{\rho}_{31}, \\
\frac{\partial\tilde{\rho}_{21}}{\partial t}=&\left(\ii\Delta_2-\Gamma_2/2\right)\tilde{\rho}_{21} -\ii\Omega_3\tilde{\rho}_{23} \\ &-\ii\left(\Omega_2+G\tilde{\rho}_{21}\right)\left(\rho_{22}-\rho_{11}\right), \nonumber \\
\frac{\partial\tilde{\rho}_{31}}{\partial t}=&\left(\ii\Delta_3-\Gamma_3/2\right)\tilde{\rho}_{31}-\ii\Omega_3\left(\rho_{33}-\rho_{11}\right)  \\
&-\ii\left(\Omega_2+G\tilde{\rho}_{21}\right)\tilde{\rho}_{32}, \nonumber \\
\frac{\partial\tilde{\rho}_{23}}{\partial t}=&\left[\ii\left(\Delta_2-\Delta_3\right)-\left(\Gamma_2+\Gamma_3\right)/2\right]\tilde{\rho}_{23}  \\
& - \ii\Omega^*_3\tilde{\rho}_{21}+\ii\left(\Omega_2+G\tilde{\rho}_{21}\right)\tilde{\rho}_{13},  \nonumber
\end{align}
\end{subequations}
where $\Omega_2=\Omega$ [see Eq.\ (\ref{O})], $\Omega_3=\vmu_{13}\cdot\Eb_{\rm probe}/\hbar$, $\Delta_2\equiv\sw-\varepsilon_{12}$, $\Delta_3\equiv\omega_{\rm probe}-\varepsilon_{13}$, and we have transformed the coherence elements according to
\begin{subequations}
\begin{align}
\tilde{\rho}_{21}&=\rho_{21}\ee^{-\ii\sw t}, \\
\tilde{\rho}_{31}&=\rho_{31}\ee^{-\ii\omega_{\rm probe}t}, \\
\tilde{\rho}_{23}&=\rho_{23}\ee^{-\ii(\sw-\omega_{\rm probe})t}.
\end{align}
\end{subequations}
In practical calculations, we consider a resonant driving nonlinear near-field (i.e., $\Delta_2=0$) with a weakly-driving probe field, such that $\Omega_3=0.01\Gamma_0$, while setting $\Gamma_2=\Gamma_3=\Gamma_0$ for simplicity. Electromagnetically-induced transparency is observed in the probe field absorption, which is proportional to ${\rm Im}\{\tilde{\rho}_{31}\}$, as the detuning $\Delta_3$ is varied.

\subsection{Three-level $\Lambda$-type atom: Temporal quantum control}

A $\Lambda$-type atom is characterized by two low-energy, nearly-degenerate states $\ket{1}$ and $\ket{2}$ that are each optically coupled to a higher-energy excited state $\ket{3}$. We study the interaction of the atom with a classical pulsed laser field
\begin{equation}
\Eb(\rb,t)=\Eb^{ns}_{\rm ind}(\rb,\omega)\ee^{-\ii\sw t}\ee^{-s(t-t_{\rm p})^2/2\tau_{\rm p}^2}+{\rm c.c.},
\end{equation}
where $t_{\rm p}$ is the temporal center of the impinging pulse and $\tau_{\rm p}$ quantifies its duration, which is considered to be much shorter than the QE excited state lifetime $\Gamma_0^{-1}$. We define the energy half-splitting between states $\ket{1}$ and $\ket{2}$ as $\Delta\equiv(\varepsilon_2-\varepsilon_1)/2$ and assume a scenario in which the pulse carrier frequency is centered between the transitions $\ket{1}\leftrightarrow\ket{3}$ and $\ket{2}\leftrightarrow\ket{3}$, such that $\sw=\varepsilon_{13}-\Delta=\varepsilon_{23}+\Delta$. The system Hamiltonian is then written in the rotating-wave approximation as
\begin{align}
\mathcal{H}=&\hbar\sum^3_{j=1}\varepsilon_j\sigma_{jj}-\ee^{-s(t-t_{\rm p})^2/2\tau_{\rm p}^2}  \\
&\times\left[\left(\vmu_{13}\sigma_{31}+\vmu_{23}\sigma_{32}\right)\cdot\Eb^{ns}_{\rm ind}(\rb,\omega)\ee^{-\ii\sw t}+{\rm h.c.}\right],  \nonumber
\end{align}
while we write the Lindblad operator as
\begin{equation}
\mathcal{L}[\rho]=\sum^2_{j=1}\frac{\Gamma_j}{2}\left(2\sigma_{j3}\rho\sigma_{3j}-\sigma_{3j}\sigma_{j3}\rho-\rho\sigma_{3j}\sigma_{j3}\right),
\end{equation}
where $\Gamma_j$ denotes the decay rate of the excited state $\ket{3}$ to state $\ket{j}$. Assuming that the pulse bandwidth is much larger than the energy splitting $2\Delta$, we use the above expressions in Eq.\ (\ref{master}) to write the density matrix equations of motion as
\begin{subequations}
\begin{align}
\frac{\partial\rho_{11}}{\partial t}=&\Gamma_1\rho_{33}+\ii\left(\Omega^*_1+G_1^*\tilde{\rho}_{13}\right)\tilde{\rho}_{31}-\ii\left(\Omega_1+G_1\tilde{\rho}_{31}\right)\tilde{\rho}_{13}, \\
\frac{\partial\rho_{22}}{\partial t}=&\Gamma_2\rho_{33}+\ii\left(\Omega^*_2+G_2^*\tilde{\rho}_{23}\right)\tilde{\rho}_{32}-\ii\left(\Omega_2+G_2\tilde{\rho}_{32}\right)\tilde{\rho}_{23}, \\
\frac{\partial{\rho}_{12}}{\partial t}=&2\ii\Delta{\rho}_{12}+\ii\left(\Omega^*_1+G_1^*\tilde{\rho}_{13}\right)\tilde{\rho}_{32}-\ii\left(\Omega_2+G_2\tilde{\rho}_{32}\right)\tilde{\rho}_{13}, \\
\frac{\partial\tilde{\rho}_{31}}{\partial t}=&-\left[\left(\Gamma_1+\Gamma_2\right)/2+\ii\Delta\right]\tilde{\rho}_{31}+\ii\left(\Omega_2+G_2\tilde{\rho}_{32}\right){\rho}_{21}  \nonumber \\
&+\ii\left(\Omega_1+G_1\tilde{\rho}_{31}\right)\left(\rho_{11}-\rho_{33}\right) ,   \\
\frac{\partial\tilde{\rho}_{32}}{\partial t}=&-\left[\left(\Gamma_1+\Gamma_2\right)/2-\ii\Delta\right]\tilde{\rho}_{32}+\ii\left(\Omega_1+G_1\tilde{\rho}_{31}\right){\rho}_{12}  \nonumber \\
&+\ii\left(\Omega_2+G_2\tilde{\rho}_{32}\right)\left(\rho_{11}-\rho_{22}\right) . 
\end{align}
\end{subequations}
In the above expressions, we have defined
\begin{equation}
\Omega_j=\frac{\ii\ee^{-s(t-t_{\rm p})^2/2\tau_{\rm p}^2}}{s\hbar\omega_j\epsa_{s\omega_j} D}\sum_m b^{ns}_m(\omega_j)\frac{\vmu_{j3}\cdot\eb_m}{1-\eta^{(1)}_{s\omega_j}/\eta_m}
\end{equation}
and
\begin{equation}
G_j=\frac{\ii\sigma^{(1)}_{\sw_j}}{s\hbar\omega_j(\epsa_{s\omega_j})^2 D^4}\sum_m\frac{\left(\vmu_{j3}\cdot\eb_m\right)^2}{1-\eta^{(1)}_{s\omega_j}/\eta_m},
\end{equation}
where $s\omega_j=\varepsilon_{j3}$, and we have transformed the coherence elements according to $\tilde{\rho}_{j3}=\rho_{j3}\ee^{-\ii s\omega t}$ ($j\neq3$). For the calculations presented in the main text, we take the transition dipole moments $\vmu_{13}$ and $\vmu_{23}$ to be orthogonal and each oriented $45^{\circ}$ degrees from the impinging field polarization, so that in the absence of material dispersion (e.g., without plasmonic enhancement) each of the transitions would be driven equally. For simplicity we set all decay rates equal to $\Gamma_0$.


\begin{thebibliography}{49}
\expandafter\ifx\csname natexlab\endcsname\relax\def\natexlab#1{#1}\fi
\expandafter\ifx\csname bibnamefont\endcsname\relax
  \def\bibnamefont#1{#1}\fi
\expandafter\ifx\csname bibfnamefont\endcsname\relax
  \def\bibfnamefont#1{#1}\fi
\expandafter\ifx\csname citenamefont\endcsname\relax
  \def\citenamefont#1{#1}\fi
\expandafter\ifx\csname url\endcsname\relax
  \def\url#1{\texttt{#1}}\fi
\expandafter\ifx\csname urlprefix\endcsname\relax\def\urlprefix{URL }\fi
\providecommand{\bibinfo}[2]{#2}
\providecommand{\eprint}[2][]{\url{#2}}

\bibitem[{\citenamefont{Atwater and Polman}(2010)}]{AP10}
\bibinfo{author}{\bibfnamefont{H.~A.} \bibnamefont{Atwater}} \bibnamefont{and}
  \bibinfo{author}{\bibfnamefont{A.}~\bibnamefont{Polman}},
  \bibinfo{journal}{Nat.\ Mater.} \textbf{\bibinfo{volume}{9}},
  \bibinfo{pages}{205} (\bibinfo{year}{2010}).

\bibitem[{\citenamefont{Fabrizio et~al.}(2016)\citenamefont{Fabrizio,
  Schl\"{u}cker, Wenger, Regmi, Rigneault, Calafiore, West, Cabrini, Fleischer,
  van Hulst et~al.}}]{DSW16}
\bibinfo{author}{\bibfnamefont{E.~D.} \bibnamefont{Fabrizio}},
  \bibinfo{author}{\bibfnamefont{S.}~\bibnamefont{Schl\"{u}cker}},
  \bibinfo{author}{\bibfnamefont{J.}~\bibnamefont{Wenger}},
  \bibinfo{author}{\bibfnamefont{R.}~\bibnamefont{Regmi}},
  \bibinfo{author}{\bibfnamefont{H.}~\bibnamefont{Rigneault}},
  \bibinfo{author}{\bibfnamefont{G.}~\bibnamefont{Calafiore}},
  \bibinfo{author}{\bibfnamefont{M.}~\bibnamefont{West}},
  \bibinfo{author}{\bibfnamefont{S.}~\bibnamefont{Cabrini}},
  \bibinfo{author}{\bibfnamefont{M.}~\bibnamefont{Fleischer}},
  \bibinfo{author}{\bibfnamefont{N.~F.} \bibnamefont{van Hulst}},
  \bibnamefont{et~al.}, \bibinfo{journal}{J.\ Opt.}
  \textbf{\bibinfo{volume}{18}}, \bibinfo{pages}{063003}
  (\bibinfo{year}{2016}).

\bibitem[{\citenamefont{Lodahl et~al.}(2015)\citenamefont{Lodahl, Mahmoodian,
  and Stobbe}}]{LMS15}
\bibinfo{author}{\bibfnamefont{P.}~\bibnamefont{Lodahl}},
  \bibinfo{author}{\bibfnamefont{S.}~\bibnamefont{Mahmoodian}},
  \bibnamefont{and} \bibinfo{author}{\bibfnamefont{S.}~\bibnamefont{Stobbe}},
  \bibinfo{journal}{Rev.\ Mod.\ Phys.} \textbf{\bibinfo{volume}{87}},
  \bibinfo{pages}{347} (\bibinfo{year}{2015}).

\bibitem[{\citenamefont{Maier}(2007)}]{M07}
\bibinfo{author}{\bibfnamefont{S.~A.} \bibnamefont{Maier}},
  \emph{\bibinfo{title}{Plasmonics: Fundamentals and Applications}}
  (\bibinfo{publisher}{Springer}, \bibinfo{address}{New York},
  \bibinfo{year}{2007}).

\bibitem[{\citenamefont{Schuller et~al.}(2010)\citenamefont{Schuller, Barnard,
  Cai, Jun, White, and Brongersma}}]{SBC10}
\bibinfo{author}{\bibfnamefont{J.~A.} \bibnamefont{Schuller}},
  \bibinfo{author}{\bibfnamefont{E.~S.} \bibnamefont{Barnard}},
  \bibinfo{author}{\bibfnamefont{W.}~\bibnamefont{Cai}},
  \bibinfo{author}{\bibfnamefont{Y.~C.} \bibnamefont{Jun}},
  \bibinfo{author}{\bibfnamefont{J.~S.} \bibnamefont{White}}, \bibnamefont{and}
  \bibinfo{author}{\bibfnamefont{M.~L.} \bibnamefont{Brongersma}},
  \bibinfo{journal}{Nat.\ Mater.} \textbf{\bibinfo{volume}{9}},
  \bibinfo{pages}{193} (\bibinfo{year}{2010}).

\bibitem[{\citenamefont{Achermann}(2010)}]{A10}
\bibinfo{author}{\bibfnamefont{M.}~\bibnamefont{Achermann}},
  \bibinfo{journal}{J.\ Phys.\ Chem.\ Lett.} \textbf{\bibinfo{volume}{1}},
  \bibinfo{pages}{2837} (\bibinfo{year}{2010}).

\bibitem[{\citenamefont{Brolo}(2012)}]{B12}
\bibinfo{author}{\bibfnamefont{A.~G.} \bibnamefont{Brolo}},
  \bibinfo{journal}{Nat.\ Photon.} \textbf{\bibinfo{volume}{6}},
  \bibinfo{pages}{709} (\bibinfo{year}{2012}).

\bibitem[{\citenamefont{Koenderink}(2017)}]{K17}
\bibinfo{author}{\bibfnamefont{A.~F.} \bibnamefont{Koenderink}},
  \bibinfo{journal}{ACS\ Photon.} \textbf{\bibinfo{volume}{4}},
  \bibinfo{pages}{710} (\bibinfo{year}{2017}).

\bibitem[{\citenamefont{Brown et~al.}(2015)\citenamefont{Brown, Sheldon, and
  Atwater}}]{BSA15}
\bibinfo{author}{\bibfnamefont{A.~M.} \bibnamefont{Brown}},
  \bibinfo{author}{\bibfnamefont{M.~T.} \bibnamefont{Sheldon}},
  \bibnamefont{and} \bibinfo{author}{\bibfnamefont{H.~A.}
  \bibnamefont{Atwater}}, \bibinfo{journal}{ACS\ Photon.}
  \textbf{\bibinfo{volume}{2}}, \bibinfo{pages}{459} (\bibinfo{year}{2015}).

\bibitem[{\citenamefont{Khurgin}(2015)}]{K15_2}
\bibinfo{author}{\bibfnamefont{J.~B.} \bibnamefont{Khurgin}},
  \bibinfo{journal}{Nat.\ Nanotech.} \textbf{\bibinfo{volume}{10}},
  \bibinfo{pages}{2} (\bibinfo{year}{2015}).

\bibitem[{\citenamefont{{Castro Neto} et~al.}(2009)\citenamefont{{Castro Neto},
  Guinea, Peres, Novoselov, and Geim}}]{CGP09}
\bibinfo{author}{\bibfnamefont{A.~H.} \bibnamefont{{Castro Neto}}},
  \bibinfo{author}{\bibfnamefont{F.}~\bibnamefont{Guinea}},
  \bibinfo{author}{\bibfnamefont{N.~M.~R.} \bibnamefont{Peres}},
  \bibinfo{author}{\bibfnamefont{K.~S.} \bibnamefont{Novoselov}},
  \bibnamefont{and} \bibinfo{author}{\bibfnamefont{A.~K.} \bibnamefont{Geim}},
  \bibinfo{journal}{Rev.\ Mod.\ Phys.} \textbf{\bibinfo{volume}{81}},
  \bibinfo{pages}{109} (\bibinfo{year}{2009}).

\bibitem[{\citenamefont{Grigorenko et~al.}(2012)\citenamefont{Grigorenko,
  Polini, and Novoselov}}]{GPN12}
\bibinfo{author}{\bibfnamefont{A.~N.} \bibnamefont{Grigorenko}},
  \bibinfo{author}{\bibfnamefont{M.}~\bibnamefont{Polini}}, \bibnamefont{and}
  \bibinfo{author}{\bibfnamefont{K.~S.} \bibnamefont{Novoselov}},
  \bibinfo{journal}{Nat.\ Photon.} \textbf{\bibinfo{volume}{6}},
  \bibinfo{pages}{749} (\bibinfo{year}{2012}).

\bibitem[{\citenamefont{Koppens et~al.}(2011)\citenamefont{Koppens, Chang, and
  {Garc\'{\i}a de Abajo}}}]{paper176}
\bibinfo{author}{\bibfnamefont{F.~H.~L.} \bibnamefont{Koppens}},
  \bibinfo{author}{\bibfnamefont{D.~E.} \bibnamefont{Chang}}, \bibnamefont{and}
  \bibinfo{author}{\bibfnamefont{F.~J.} \bibnamefont{{Garc\'{\i}a de Abajo}}},
  \bibinfo{journal}{Nano\ Lett.} \textbf{\bibinfo{volume}{11}},
  \bibinfo{pages}{3370} (\bibinfo{year}{2011}).

\bibitem[{\citenamefont{Nikitin et~al.}(2011)\citenamefont{Nikitin, Guinea,
  Garc\'{\i}a-Vidal, and Mart\'{\i}n-Moreno}}]{NGG11_2}
\bibinfo{author}{\bibfnamefont{A.~Y.} \bibnamefont{Nikitin}},
  \bibinfo{author}{\bibfnamefont{F.}~\bibnamefont{Guinea}},
  \bibinfo{author}{\bibfnamefont{F.~J.} \bibnamefont{Garc\'{\i}a-Vidal}},
  \bibnamefont{and}
  \bibinfo{author}{\bibfnamefont{L.}~\bibnamefont{Mart\'{\i}n-Moreno}},
  \bibinfo{journal}{Phys.\ Rev.\ B} \textbf{\bibinfo{volume}{84}},
  \bibinfo{pages}{195446} (\bibinfo{year}{2011}).

\bibitem[{\citenamefont{Sloan et~al.}(2018)\citenamefont{Sloan, Rivera,
  Solja{\v{c}}i{\'c}, and Kaminer}}]{SRS18}
\bibinfo{author}{\bibfnamefont{J.}~\bibnamefont{Sloan}},
  \bibinfo{author}{\bibfnamefont{N.}~\bibnamefont{Rivera}},
  \bibinfo{author}{\bibfnamefont{M.}~\bibnamefont{Solja{\v{c}}i{\'c}}},
  \bibnamefont{and} \bibinfo{author}{\bibfnamefont{I.}~\bibnamefont{Kaminer}},
  \bibinfo{journal}{Nano\ Lett.} \textbf{\bibinfo{volume}{18}},
  \bibinfo{pages}{308} (\bibinfo{year}{2018}).

\bibitem[{\citenamefont{Manjavacas et~al.}(2012)\citenamefont{Manjavacas,
  Thongrattanasiri, Chang, and {Garc\'{\i}a de Abajo}}}]{paper204}
\bibinfo{author}{\bibfnamefont{A.}~\bibnamefont{Manjavacas}},
  \bibinfo{author}{\bibfnamefont{S.}~\bibnamefont{Thongrattanasiri}},
  \bibinfo{author}{\bibfnamefont{D.~E.} \bibnamefont{Chang}}, \bibnamefont{and}
  \bibinfo{author}{\bibfnamefont{F.~J.} \bibnamefont{{Garc\'{\i}a de Abajo}}},
  \bibinfo{journal}{New\ J.\ Phys.} \textbf{\bibinfo{volume}{14}},
  \bibinfo{pages}{123020} (\bibinfo{year}{2012}).

\bibitem[{\citenamefont{Yan et~al.}(2013)\citenamefont{Yan, Low, Zhu, Wu,
  Freitag, Li, Guinea, Avouris, and Xia}}]{YLZ12}
\bibinfo{author}{\bibfnamefont{H.}~\bibnamefont{Yan}},
  \bibinfo{author}{\bibfnamefont{T.}~\bibnamefont{Low}},
  \bibinfo{author}{\bibfnamefont{W.}~\bibnamefont{Zhu}},
  \bibinfo{author}{\bibfnamefont{Y.}~\bibnamefont{Wu}},
  \bibinfo{author}{\bibfnamefont{M.}~\bibnamefont{Freitag}},
  \bibinfo{author}{\bibfnamefont{X.}~\bibnamefont{Li}},
  \bibinfo{author}{\bibfnamefont{F.}~\bibnamefont{Guinea}},
  \bibinfo{author}{\bibfnamefont{P.}~\bibnamefont{Avouris}}, \bibnamefont{and}
  \bibinfo{author}{\bibfnamefont{F.}~\bibnamefont{Xia}},
  \bibinfo{journal}{Nat.\ Photon.} \textbf{\bibinfo{volume}{7}},
  \bibinfo{pages}{394} (\bibinfo{year}{2013}).

\bibitem[{\citenamefont{{Garc\'{\i}a de Abajo}}(2014)}]{paper235}
\bibinfo{author}{\bibfnamefont{F.~J.} \bibnamefont{{Garc\'{\i}a de Abajo}}},
  \bibinfo{journal}{ACS\ Photon.} \textbf{\bibinfo{volume}{1}},
  \bibinfo{pages}{135} (\bibinfo{year}{2014}).

\bibitem[{\citenamefont{Mikhailov}(2007)}]{M07_2}
\bibinfo{author}{\bibfnamefont{S.~A.} \bibnamefont{Mikhailov}},
  \bibinfo{journal}{Europhys.\ Lett.} \textbf{\bibinfo{volume}{79}},
  \bibinfo{pages}{27002} (\bibinfo{year}{2007}).

\bibitem[{\citenamefont{Ishikawa}(2010)}]{I10}
\bibinfo{author}{\bibfnamefont{K.~L.} \bibnamefont{Ishikawa}},
  \bibinfo{journal}{Phys.\ Rev.\ B} \textbf{\bibinfo{volume}{82}},
  \bibinfo{pages}{201402(R)} (\bibinfo{year}{2010}).

\bibitem[{\citenamefont{Cheng et~al.}(2015)\citenamefont{Cheng, Vermeulen, and
  Sipe}}]{CVS15}
\bibinfo{author}{\bibfnamefont{J.~L.} \bibnamefont{Cheng}},
  \bibinfo{author}{\bibfnamefont{N.}~\bibnamefont{Vermeulen}},
  \bibnamefont{and} \bibinfo{author}{\bibfnamefont{J.~E.} \bibnamefont{Sipe}},
  \bibinfo{journal}{Phys.\ Rev.\ B} \textbf{\bibinfo{volume}{91}},
  \bibinfo{pages}{235320} (\bibinfo{year}{2015}).

\bibitem[{\citenamefont{Mikhailov}(2016)}]{M16}
\bibinfo{author}{\bibfnamefont{S.~A.} \bibnamefont{Mikhailov}},
  \bibinfo{journal}{Phys.\ Rev.\ B} \textbf{\bibinfo{volume}{93}},
  \bibinfo{pages}{085403} (\bibinfo{year}{2016}).

\bibitem[{\citenamefont{Jiang et~al.}(2018)\citenamefont{Jiang, Huang, Cheng,
  Fan, Zhang, Shan, Yi, Dai, Shi, Liu et~al.}}]{JHC18}
\bibinfo{author}{\bibfnamefont{T.}~\bibnamefont{Jiang}},
  \bibinfo{author}{\bibfnamefont{D.}~\bibnamefont{Huang}},
  \bibinfo{author}{\bibfnamefont{J.}~\bibnamefont{Cheng}},
  \bibinfo{author}{\bibfnamefont{X.}~\bibnamefont{Fan}},
  \bibinfo{author}{\bibfnamefont{Z.}~\bibnamefont{Zhang}},
  \bibinfo{author}{\bibfnamefont{Y.}~\bibnamefont{Shan}},
  \bibinfo{author}{\bibfnamefont{Y.}~\bibnamefont{Yi}},
  \bibinfo{author}{\bibfnamefont{Y.}~\bibnamefont{Dai}},
  \bibinfo{author}{\bibfnamefont{L.}~\bibnamefont{Shi}},
  \bibinfo{author}{\bibfnamefont{K.}~\bibnamefont{Liu}}, \bibnamefont{et~al.},
  \bibinfo{journal}{Nat.\ Photon.} \textbf{\bibinfo{volume}{12}},
  \bibinfo{pages}{430} (\bibinfo{year}{2018}).

\bibitem[{\citenamefont{Soavil et~al.}(2018)\citenamefont{Soavil, Wang,
  Rostami, Purdie, Fazio, Ma, Luo, Wang, Ott, Yoon et~al.}}]{SWR18}
\bibinfo{author}{\bibfnamefont{G.}~\bibnamefont{Soavil}},
  \bibinfo{author}{\bibfnamefont{G.}~\bibnamefont{Wang}},
  \bibinfo{author}{\bibfnamefont{H.}~\bibnamefont{Rostami}},
  \bibinfo{author}{\bibfnamefont{D.~G.} \bibnamefont{Purdie}},
  \bibinfo{author}{\bibfnamefont{D.~D.} \bibnamefont{Fazio}},
  \bibinfo{author}{\bibfnamefont{T.}~\bibnamefont{Ma}},
  \bibinfo{author}{\bibfnamefont{B.}~\bibnamefont{Luo}},
  \bibinfo{author}{\bibfnamefont{J.}~\bibnamefont{Wang}},
  \bibinfo{author}{\bibfnamefont{A.~K.} \bibnamefont{Ott}},
  \bibinfo{author}{\bibfnamefont{D.}~\bibnamefont{Yoon}}, \bibnamefont{et~al.},
  \bibinfo{journal}{Nat.\ Nanotech.} \textbf{\bibinfo{volume}{13}},
  \bibinfo{pages}{583} (\bibinfo{year}{2018}).

\bibitem[{\citenamefont{Hafez et~al.}(2018)\citenamefont{Hafez, Kovalev,
  Deinert, Mics, Green, Awari, Chen, Germanskiy, Lehnert, Teichert
  et~al.}}]{HKD18}
\bibinfo{author}{\bibfnamefont{H.~A.} \bibnamefont{Hafez}},
  \bibinfo{author}{\bibfnamefont{S.}~\bibnamefont{Kovalev}},
  \bibinfo{author}{\bibfnamefont{J.-C.} \bibnamefont{Deinert}},
  \bibinfo{author}{\bibfnamefont{Z.}~\bibnamefont{Mics}},
  \bibinfo{author}{\bibfnamefont{B.}~\bibnamefont{Green}},
  \bibinfo{author}{\bibfnamefont{N.}~\bibnamefont{Awari}},
  \bibinfo{author}{\bibfnamefont{M.}~\bibnamefont{Chen}},
  \bibinfo{author}{\bibfnamefont{S.}~\bibnamefont{Germanskiy}},
  \bibinfo{author}{\bibfnamefont{U.}~\bibnamefont{Lehnert}},
  \bibinfo{author}{\bibfnamefont{J.}~\bibnamefont{Teichert}},
  \bibnamefont{et~al.}, \bibinfo{journal}{Nature}
  \textbf{\bibinfo{volume}{561}}, \bibinfo{pages}{507} (\bibinfo{year}{2018}).

\bibitem[{\citenamefont{Cox and {Garc\'{\i}a de Abajo}}(2014)}]{paper247}
\bibinfo{author}{\bibfnamefont{J.~D.} \bibnamefont{Cox}} \bibnamefont{and}
  \bibinfo{author}{\bibfnamefont{F.~J.} \bibnamefont{{Garc\'{\i}a de Abajo}}},
  \bibinfo{journal}{Nat.\ Commun.} \textbf{\bibinfo{volume}{5}},
  \bibinfo{pages}{5725} (\bibinfo{year}{2014}).

\bibitem[{\citenamefont{Jablan and Chang}(2015)}]{JC15}
\bibinfo{author}{\bibfnamefont{M.}~\bibnamefont{Jablan}} \bibnamefont{and}
  \bibinfo{author}{\bibfnamefont{D.~E.} \bibnamefont{Chang}},
  \bibinfo{journal}{Phys.\ Rev.\ Lett.} \textbf{\bibinfo{volume}{114}},
  \bibinfo{pages}{236801} (\bibinfo{year}{2015}).

\bibitem[{\citenamefont{Christensen et~al.}(2015)\citenamefont{Christensen,
  Yan, Jauho, Wubs, and Mortensen}}]{CYJ15}
\bibinfo{author}{\bibfnamefont{T.}~\bibnamefont{Christensen}},
  \bibinfo{author}{\bibfnamefont{W.}~\bibnamefont{Yan}},
  \bibinfo{author}{\bibfnamefont{A.-P.} \bibnamefont{Jauho}},
  \bibinfo{author}{\bibfnamefont{M.}~\bibnamefont{Wubs}}, \bibnamefont{and}
  \bibinfo{author}{\bibfnamefont{N.~A.} \bibnamefont{Mortensen}},
  \bibinfo{journal}{Phys.\ Rev.\ B} \textbf{\bibinfo{volume}{92}},
  \bibinfo{pages}{121407(R)} (\bibinfo{year}{2015}).

\bibitem[{\citenamefont{Constant et~al.}(2016)\citenamefont{Constant, Hornett,
  Chang, and Hendry}}]{CHC16}
\bibinfo{author}{\bibfnamefont{T.~J.} \bibnamefont{Constant}},
  \bibinfo{author}{\bibfnamefont{S.~M.} \bibnamefont{Hornett}},
  \bibinfo{author}{\bibfnamefont{D.~E.} \bibnamefont{Chang}}, \bibnamefont{and}
  \bibinfo{author}{\bibfnamefont{E.}~\bibnamefont{Hendry}},
  \bibinfo{journal}{Nat.\ Phys.} \textbf{\bibinfo{volume}{12}},
  \bibinfo{pages}{124} (\bibinfo{year}{2016}).

\bibitem[{\citenamefont{Jadidi et~al.}(2016)\citenamefont{Jadidi,
  K\"{o}nig-Otto, Winnerl, Sushkov, Drew, Murphy, and Mittendorff}}]{JKW16}
\bibinfo{author}{\bibfnamefont{M.~M.} \bibnamefont{Jadidi}},
  \bibinfo{author}{\bibfnamefont{J.~C.} \bibnamefont{K\"{o}nig-Otto}},
  \bibinfo{author}{\bibfnamefont{S.}~\bibnamefont{Winnerl}},
  \bibinfo{author}{\bibfnamefont{A.~B.} \bibnamefont{Sushkov}},
  \bibinfo{author}{\bibfnamefont{H.~D.} \bibnamefont{Drew}},
  \bibinfo{author}{\bibfnamefont{T.~E.} \bibnamefont{Murphy}},
  \bibnamefont{and}
  \bibinfo{author}{\bibfnamefont{M.}~\bibnamefont{Mittendorff}},
  \bibinfo{journal}{Nano\ Lett.} \textbf{\bibinfo{volume}{16}},
  \bibinfo{pages}{2734} (\bibinfo{year}{2016}).

\bibitem[{\citenamefont{Cox et~al.}(2017{\natexlab{a}})\citenamefont{Cox,
  Marini, and {Garc\'{\i}a de Abajo}}}]{paper287}
\bibinfo{author}{\bibfnamefont{J.~D.} \bibnamefont{Cox}},
  \bibinfo{author}{\bibfnamefont{A.}~\bibnamefont{Marini}}, \bibnamefont{and}
  \bibinfo{author}{\bibfnamefont{F.~J.} \bibnamefont{{Garc\'{\i}a de Abajo}}},
  \bibinfo{journal}{Nat.\ Commun.} \textbf{\bibinfo{volume}{8}},
  \bibinfo{pages}{14380} (\bibinfo{year}{2017}{\natexlab{a}}).

\bibitem[{\citenamefont{Cox et~al.}(2017{\natexlab{b}})\citenamefont{Cox, Yu,
  and {Garc\'{\i}a de Abajo}}}]{paper293}
\bibinfo{author}{\bibfnamefont{J.~D.} \bibnamefont{Cox}},
  \bibinfo{author}{\bibfnamefont{R.}~\bibnamefont{Yu}}, \bibnamefont{and}
  \bibinfo{author}{\bibfnamefont{F.~J.} \bibnamefont{{Garc\'{\i}a de Abajo}}},
  \bibinfo{journal}{Phys.\ Rev.\ B} \textbf{\bibinfo{volume}{96}},
  \bibinfo{pages}{045442} (\bibinfo{year}{2017}{\natexlab{b}}).

\bibitem[{\citenamefont{Kundys et~al.}(2018)\citenamefont{Kundys, Duppen,
  Marshall, Rodriguez, Torre, Tomadin, Polini, and Grigorenko}}]{KDM18}
\bibinfo{author}{\bibfnamefont{D.}~\bibnamefont{Kundys}},
  \bibinfo{author}{\bibfnamefont{B.~V.} \bibnamefont{Duppen}},
  \bibinfo{author}{\bibfnamefont{O.~P.} \bibnamefont{Marshall}},
  \bibinfo{author}{\bibfnamefont{F.}~\bibnamefont{Rodriguez}},
  \bibinfo{author}{\bibfnamefont{I.}~\bibnamefont{Torre}},
  \bibinfo{author}{\bibfnamefont{A.}~\bibnamefont{Tomadin}},
  \bibinfo{author}{\bibfnamefont{M.}~\bibnamefont{Polini}}, \bibnamefont{and}
  \bibinfo{author}{\bibfnamefont{A.~N.} \bibnamefont{Grigorenko}},
  \bibinfo{journal}{Nano\ Lett.} \textbf{\bibinfo{volume}{18}},
  \bibinfo{pages}{282} (\bibinfo{year}{2018}).

\bibitem[{\citenamefont{Metzger et~al.}(2017)\citenamefont{Metzger, Hentschel,
  and Giessen}}]{MHG17}
\bibinfo{author}{\bibfnamefont{B.}~\bibnamefont{Metzger}},
  \bibinfo{author}{\bibfnamefont{M.}~\bibnamefont{Hentschel}},
  \bibnamefont{and} \bibinfo{author}{\bibfnamefont{H.}~\bibnamefont{Giessen}},
  \bibinfo{journal}{Nano\ Lett.} \textbf{\bibinfo{volume}{17}},
  \bibinfo{pages}{1931} (\bibinfo{year}{2017}).

\bibitem[{\citenamefont{Yang et~al.}(2017)\citenamefont{Yang, Butet, Yan,
  Bernasconi, and Martin}}]{YBY17}
\bibinfo{author}{\bibfnamefont{K.-Y.} \bibnamefont{Yang}},
  \bibinfo{author}{\bibfnamefont{J.}~\bibnamefont{Butet}},
  \bibinfo{author}{\bibfnamefont{C.}~\bibnamefont{Yan}},
  \bibinfo{author}{\bibfnamefont{G.~D.} \bibnamefont{Bernasconi}},
  \bibnamefont{and} \bibinfo{author}{\bibfnamefont{O.~J.~F.}
  \bibnamefont{Martin}}, \bibinfo{journal}{ACS\ Photon.}
  \textbf{\bibinfo{volume}{4}}, \bibinfo{pages}{1522} (\bibinfo{year}{2017}).

\bibitem[{\citenamefont{Stievater et~al.}(2001)\citenamefont{Stievater, Li,
  Steel, Gammon, Katzer, Park, Piermarocchi, and Sham}}]{SLS01}
\bibinfo{author}{\bibfnamefont{T.~H.} \bibnamefont{Stievater}},
  \bibinfo{author}{\bibfnamefont{X.}~\bibnamefont{Li}},
  \bibinfo{author}{\bibfnamefont{D.~G.} \bibnamefont{Steel}},
  \bibinfo{author}{\bibfnamefont{D.}~\bibnamefont{Gammon}},
  \bibinfo{author}{\bibfnamefont{D.~S.} \bibnamefont{Katzer}},
  \bibinfo{author}{\bibfnamefont{D.}~\bibnamefont{Park}},
  \bibinfo{author}{\bibfnamefont{C.}~\bibnamefont{Piermarocchi}},
  \bibnamefont{and} \bibinfo{author}{\bibfnamefont{L.~J.} \bibnamefont{Sham}},
  \bibinfo{journal}{Phys.\ Rev.\ Lett.} \textbf{\bibinfo{volume}{87}},
  \bibinfo{pages}{133603} (\bibinfo{year}{2001}).

  
\bibitem[{\citenamefont{Yu et~al.}(2017{\natexlab{a}})\citenamefont{Yu, Cox,
  Saavedra, and {Garc\'{\i}a de Abajo}}}]{paper303}
\bibinfo{author}{\bibfnamefont{R.}~\bibnamefont{Yu}},
  \bibinfo{author}{\bibfnamefont{J.~D.} \bibnamefont{Cox}},
  \bibinfo{author}{\bibfnamefont{J.~R.~M.} \bibnamefont{Saavedra}},
  \bibnamefont{and} \bibinfo{author}{\bibfnamefont{F.~J.}
  \bibnamefont{{Garc\'{\i}a de Abajo}}}, \bibinfo{journal}{ACS\ Photon.}
  \textbf{\bibinfo{volume}{4}}, \bibinfo{pages}{3106}
  (\bibinfo{year}{2017}{\natexlab{a}}).

\bibitem[{\citenamefont{Wunsch et~al.}(2006)\citenamefont{Wunsch, Stauber,
  Sols, and Guinea}}]{WSS06}
\bibinfo{author}{\bibfnamefont{B.}~\bibnamefont{Wunsch}},
  \bibinfo{author}{\bibfnamefont{T.}~\bibnamefont{Stauber}},
  \bibinfo{author}{\bibfnamefont{F.}~\bibnamefont{Sols}}, \bibnamefont{and}
  \bibinfo{author}{\bibfnamefont{F.}~\bibnamefont{Guinea}},
  \bibinfo{journal}{New\ J.\ Phys.} \textbf{\bibinfo{volume}{8}},
  \bibinfo{pages}{318} (\bibinfo{year}{2006}).

\bibitem[{\citenamefont{Zhang et~al.}(2006)\citenamefont{Zhang, Govorov, and
  Bryant}}]{ZGB06}
\bibinfo{author}{\bibfnamefont{W.}~\bibnamefont{Zhang}},
  \bibinfo{author}{\bibfnamefont{A.~O.} \bibnamefont{Govorov}},
  \bibnamefont{and} \bibinfo{author}{\bibfnamefont{G.~W.}
  \bibnamefont{Bryant}}, \bibinfo{journal}{Phys. Rev. Lett.}
  \textbf{\bibinfo{volume}{97}}, \bibinfo{pages}{146804}
  (\bibinfo{year}{2006}).

\bibitem[{\citenamefont{Malyshev and Malyshev}(2011)}]{MM11}
\bibinfo{author}{\bibfnamefont{A.~V.} \bibnamefont{Malyshev}} \bibnamefont{and}
  \bibinfo{author}{\bibfnamefont{V.~A.} \bibnamefont{Malyshev}},
  \bibinfo{journal}{Phys.\ Rev.\ B} \textbf{\bibinfo{volume}{84}},
  \bibinfo{pages}{035314} (\bibinfo{year}{2011}).

\bibitem[{\citenamefont{Ant{\'o}n et~al.}(2012)\citenamefont{Ant{\'o}n,
  Carre{\~n}o, Melle, Calder{\'o}n, Cabrera-Granado, Cox, and Singh}}]{ACM12}
\bibinfo{author}{\bibfnamefont{M.~A.} \bibnamefont{Ant{\'o}n}},
  \bibinfo{author}{\bibfnamefont{F.}~\bibnamefont{Carre{\~n}o}},
  \bibinfo{author}{\bibfnamefont{S.}~\bibnamefont{Melle}},
  \bibinfo{author}{\bibfnamefont{O.~G.} \bibnamefont{Calder{\'o}n}},
  \bibinfo{author}{\bibfnamefont{E.}~\bibnamefont{Cabrera-Granado}},
  \bibinfo{author}{\bibfnamefont{J.}~\bibnamefont{Cox}}, \bibnamefont{and}
  \bibinfo{author}{\bibfnamefont{M.~R.} \bibnamefont{Singh}},
  \bibinfo{journal}{Phys.\ Rev.\ B} \textbf{\bibinfo{volume}{86}},
  \bibinfo{pages}{155305} (\bibinfo{year}{2012}).

\bibitem[{\citenamefont{Yu et~al.}(2017{\natexlab{b}})\citenamefont{Yu,
  Manjavacas, and {Garc\'{\i}a de Abajo}}}]{paper286}
\bibinfo{author}{\bibfnamefont{R.}~\bibnamefont{Yu}},
  \bibinfo{author}{\bibfnamefont{A.}~\bibnamefont{Manjavacas}},
  \bibnamefont{and} \bibinfo{author}{\bibfnamefont{F.~J.}
  \bibnamefont{{Garc\'{\i}a de Abajo}}}, \bibinfo{journal}{Nat.\ Commun.}
  \textbf{\bibinfo{volume}{8}}, \bibinfo{pages}{2}
  (\bibinfo{year}{2017}{\natexlab{b}}).

\bibitem[{\citenamefont{Doeleman et~al.}(2016)\citenamefont{Doeleman, Verhagen,
  and Koenderink}}]{DVK16}
\bibinfo{author}{\bibfnamefont{H.~M.} \bibnamefont{Doeleman}},
  \bibinfo{author}{\bibfnamefont{E.}~\bibnamefont{Verhagen}}, \bibnamefont{and}
  \bibinfo{author}{\bibfnamefont{A.~F.} \bibnamefont{Koenderink}},
  \bibinfo{journal}{ACS\ Photon.} \textbf{\bibinfo{volume}{3}},
  \bibinfo{pages}{1943} (\bibinfo{year}{2016}).

\bibitem[{\citenamefont{Boyd}(2008)}]{B08_3}
\bibinfo{author}{\bibfnamefont{R.~W.} \bibnamefont{Boyd}},
  \emph{\bibinfo{title}{Nonlinear optics}} (\bibinfo{publisher}{Academic
  Press}, \bibinfo{address}{Amsterdam}, \bibinfo{year}{2008}),
  \bibinfo{edition}{3rd} ed.

\bibitem[{\citenamefont{Krauss et~al.}(2009)\citenamefont{Krauss, Lohmann,
  Chae, Haluska, von Klitzing, and Smet}}]{KLC09}
\bibinfo{author}{\bibfnamefont{B.}~\bibnamefont{Krauss}},
  \bibinfo{author}{\bibfnamefont{T.}~\bibnamefont{Lohmann}},
  \bibinfo{author}{\bibfnamefont{D.-H.} \bibnamefont{Chae}},
  \bibinfo{author}{\bibfnamefont{M.}~\bibnamefont{Haluska}},
  \bibinfo{author}{\bibfnamefont{K.}~\bibnamefont{von Klitzing}},
  \bibnamefont{and} \bibinfo{author}{\bibfnamefont{J.~H.} \bibnamefont{Smet}},
  \bibinfo{journal}{Phys.\ Rev.\ B} \textbf{\bibinfo{volume}{79}},
  \bibinfo{pages}{165428} (\bibinfo{year}{2009}).

\bibitem[{\citenamefont{Currie et~al.}(2011)\citenamefont{Currie, Caldwell,
  Bezares, Robinson, Anderson, Chun, and Tadjer}}]{CCB11}
\bibinfo{author}{\bibfnamefont{M.}~\bibnamefont{Currie}},
  \bibinfo{author}{\bibfnamefont{J.~D.} \bibnamefont{Caldwell}},
  \bibinfo{author}{\bibfnamefont{F.~J.} \bibnamefont{Bezares}},
  \bibinfo{author}{\bibfnamefont{J.}~\bibnamefont{Robinson}},
  \bibinfo{author}{\bibfnamefont{T.}~\bibnamefont{Anderson}},
  \bibinfo{author}{\bibfnamefont{H.}~\bibnamefont{Chun}}, \bibnamefont{and}
  \bibinfo{author}{\bibfnamefont{M.}~\bibnamefont{Tadjer}},
  \bibinfo{journal}{Appl.\ Phys.\ Lett.} \textbf{\bibinfo{volume}{99}},
  \bibinfo{pages}{211909} (\bibinfo{year}{2011}).

\bibitem[{\citenamefont{Roberts et~al.}(2011)\citenamefont{Roberts, Cormode,
  Reynolds, Newhouse-Illige, LeRoy, and Sandhu}}]{RCR11}
\bibinfo{author}{\bibfnamefont{A.}~\bibnamefont{Roberts}},
  \bibinfo{author}{\bibfnamefont{D.}~\bibnamefont{Cormode}},
  \bibinfo{author}{\bibfnamefont{C.}~\bibnamefont{Reynolds}},
  \bibinfo{author}{\bibfnamefont{T.}~\bibnamefont{Newhouse-Illige}},
  \bibinfo{author}{\bibfnamefont{B.~J.} \bibnamefont{LeRoy}}, \bibnamefont{and}
  \bibinfo{author}{\bibfnamefont{A.~S.} \bibnamefont{Sandhu}},
  \bibinfo{journal}{Appl.\ Phys.\ Lett.} \textbf{\bibinfo{volume}{99}},
  \bibinfo{pages}{051912} (\bibinfo{year}{2011}).

\bibitem[{\citenamefont{Kiisk et~al.}(2013)\citenamefont{Kiisk, Kahro, Kozlova,
  Matisen, and Alles}}]{KKK13}
\bibinfo{author}{\bibfnamefont{V.}~\bibnamefont{Kiisk}},
  \bibinfo{author}{\bibfnamefont{T.}~\bibnamefont{Kahro}},
  \bibinfo{author}{\bibfnamefont{J.}~\bibnamefont{Kozlova}},
  \bibinfo{author}{\bibfnamefont{L.}~\bibnamefont{Matisen}}, \bibnamefont{and}
  \bibinfo{author}{\bibfnamefont{H.}~\bibnamefont{Alles}},
  \bibinfo{journal}{Appl.\ Surf.\ Sci.} \textbf{\bibinfo{volume}{276}},
  \bibinfo{pages}{133} (\bibinfo{year}{2013}).

\bibitem[{\citenamefont{Fleischhauer et~al.}(2005)\citenamefont{Fleischhauer,
  Imamo{\u{g}}lu, and Marangos}}]{FIM05}
\bibinfo{author}{\bibfnamefont{M.}~\bibnamefont{Fleischhauer}},
  \bibinfo{author}{\bibfnamefont{A.}~\bibnamefont{Imamo{\u{g}}lu}},
  \bibnamefont{and} \bibinfo{author}{\bibfnamefont{J.~P.}
  \bibnamefont{Marangos}}, \bibinfo{journal}{Rev.\ Mod.\ Phys.}
  \textbf{\bibinfo{volume}{77}}, \bibinfo{pages}{633} (\bibinfo{year}{2005}).

\end{thebibliography}

\end{document}